\documentclass[aps,prl,twocolumn,showpacs,10pt,preprintnumbers,nofootinbib]{revtex4-1}

\usepackage[margin=7pt,justification=centerlast]{caption}

\usepackage{amsmath}
\usepackage{amsfonts}
\usepackage{amssymb}
\usepackage{graphicx, rotating}
\usepackage{epstopdf}
\usepackage{epsfig}
\usepackage{latexsym}
\usepackage{mathtools}
\usepackage{MnSymbol}
\usepackage{graphicx}
\usepackage{color}
\usepackage[dvipsnames]{xcolor}
\usepackage{amsmath,amssymb}
\usepackage{multirow}
\usepackage{hhline}
\usepackage{array,makecell}
\usepackage[utf8]{inputenc}
\usepackage{arydshln}
\usepackage{ mathrsfs }
\usepackage{orcidlink}
\usepackage{braket}
\usepackage{contour}
\usepackage{booktabs}
\usepackage{comment}

\usepackage{tikz-feynman}
\usetikzlibrary{decorations.text}
\tikzfeynmanset{warn luatex=false}

\usepackage{slashed}
\usepackage{hyperref}
\hypersetup{colorlinks, citecolor=black, linkcolor=black, urlcolor=bluscuro}
\definecolor{rossos}{cmyk}{0,1,1,0.55}
\definecolor{bluscuro}{rgb}{0.15, 0.2, .85}
\definecolor{bluchiaro}{cmyk}{1,.3,0.,0.1}
\definecolor{verdescuro}{rgb}{0.3,0.8,0.3}

\usepackage{eso-pic}
\newcommand{\reportnum}[2]{
  \AddToShipoutPictureBG*{%
    \AtPageUpperLeft{%
      \hspace{0.80\paperwidth}%
      \raisebox{#1\baselineskip}{%
        \makebox[0pt][l]{\textnormal{#2}}
  }}}%
}

\newcommand{\nn}{\nonumber}

\newcommand{\be}{\begin{equation}}
\newcommand{\ee}{\end{equation}}          
\newcommand{\bea}{\begin{eqnarray}}
\newcommand{\eea}{\end{eqnarray}}
\newcommand{\bc}{\begin{center}}
	\newcommand{\ec}{\end{center}}

\newcommand{\TeV}{\,\mathrm{TeV}}

\reportnum{-3}{CERN-TH-2023-231}

\begin{document}

\title{Boundaries of Universal Theories}

\author{Matthew McCullough\orcidlink{0000-0003-3971-9306}}
\affiliation{Theoretical Physics Department, CERN, 1211 Geneva 23, Switzerland}
\author{Lorenzo Ricci\orcidlink{0000-0001-8704-3545}}
\affiliation{Maryland Center for Fundamental Physics, Department of Physics,
	University of Maryland, College Park, MD 20742, USA}
\author{Marc Riembau\orcidlink{0000-0002-9842-2425}}
\affiliation{Theoretical Physics Department, CERN, 1211 Geneva 23, Switzerland}

\begin{abstract}
\noindent 
{
Universal theories are a broad class of well-motivated microscopic dynamics of the electroweak sector that go beyond the Standard Model description. The long distance physics is described by electroweak parameters which correspond to local operators in the Effective Field Theory.
We show how unitarity and analyticity constrain the space of parameters. In particular, the $W$ and $Y$ parameters are constrained to be positive and are necessarily the leading terms in the low energy expansion. We assess the impact of unitarity on the interpretation of Drell-Yan data.  In passing, we uncover an unexpected Wilson coefficient transcendental cancellation at the $\mathcal{O}(<10^{-3})$ level.
}

\end{abstract}

\maketitle

\medskip

\section{Introduction}
As we approach the era of the High Luminosity LHC (HL-LHC), attention will increasingly turn towards high precision probes of new physics.  One particularly powerful class of precision new physics observables are the $W$ and $Y$ electroweak (EW) parameters.  The future opportunities for effectively probing new physics through these observables at the LHC were emphasised already some time ago in \cite{Farina:2016rws}, whose central point will only become more relevant as time go on.  Recent estimates suggest that between now ($140\text{ fb}^{-1}$) and the full HL-LHC dataset ($3\text{ ab}^{-1}$) the precision on measurements of these parameters will improve roughly by a factor $2$ \cite{Torre:2020aiz,Panico:2021vav}. However, given the encouraging results of a recent CMS analysis \cite{CMS:2022krd} with $101 \text{ fb}^{-1}$ (already overcoming the $300 \text{fb}^{-1}$ projections \cite{Farina:2016rws, Torre:2020aiz}) together with new possible dedicated measurements \cite{Panico:2021vav} and continuous theoretical and experimental effort \cite{Amoroso:2022eow,Campbell:2022qmc}, there is reason to believe that the factor 2 improvement expected for HL-LHC may be overly conservative.

It is thus increasingly important to better understand how to interpret these EW parameters and any constraints on them or, better, any emerging evidence for non-zero values.  In this work we seek to further that understanding.  It is well known that these EW parameters only find a robust IR interpretation within the \emph{universal} universality class of UV completions to the Standard Model \cite{Barbieri:2004qk,Wells:2015uba}.  I.e.\ those BSM scenarios in which the \emph{leading} effects of new physics may be encoded in modified gauge boson propagators.

Theoretical constraints on the oblique parameters have been widely investigated in the past \cite{EINHORN1981146,Peskin:1990zt,Sundrum:1991rf,Barbieri:2003pr,Agashe:2007mc,PhysRevD.46.381,He:2001tp,Cacciapaglia:2006pk}. 
Here we continue this program and we show that within the class of universal theories the $W$ and $Y$ parameters are positive, under mild assumptions concerning perturbativity of the EW gauge couplings. 

This observation has significant implications for the interpretation of experimental analyses.  The present situation, in which the CMS analysis \cite{CMS:2022krd} presently observes a negative $W$ parameter at greater than the $2\sigma$ level, provides a timely opportunity to highlight those implications.

The first is that if this deviation were to persist in the future, and grow in significance, it would imply the new physics is not universal, providing crucial information on the structure of new physics in the UV.  The second implication is that, interpreted as a bound on the mass scale of new physics within the context of UV-completions where the $W$ parameter is relevant, positivity implies much stronger constraints on the mass scale of new physics than if positivity is disregarded.  In this latter regard we find a constraint on the mass scale of new physics almost a factor $3$ stronger than reported by CMS \cite{CMS:2022krd}.

This letter is structured as follows. In Sec.~\ref{sec:positivity} positivity is derived in detail.  In Sec.~\ref{sec:SM} the interpretation of universal theories is expounded by analogy with the SM, wherein two examples of universal ``UV completions'' which modify the photon propagator are studied, namely QCD and Hypercharge.  In the course of these calculations we reveal a surprising transcendental cancellation in a Wilson coefficient, for which we lack a satisfactory understanding.  In Sec.~\ref{sec:BSM} we look beyond the Standard Model and study the implications of positivity and unitarity for EW precision parameters.  In Sec.~\ref{sec:pheno} we consider the current phenomenological situation, before concluding in Sec.~\ref{sec:conc}.

\section{Positivity of oblique corrections}
\label{sec:positivity}

Consider a theory $\mathcal{T}_1$ that describes the dynamics of a system at low energy.
For simplicity and relevance for our discussion, this theory has a weakly coupled Lagrangian description and, in fact, through this paper will consist of free or weakly gauged matter. 
A different sector $\mathcal{T}_2$ may instead have a mass gap, $m_\star$, and may or may not be strongly coupled.
The sole connection between both sectors is the gauging of some symmetry $\mathcal{G}$, subgroup of the global symmetries of both $\mathcal{T}_1$ and $\mathcal{T}_2$.
A cartoon depicting this situation is:

\medskip
\centerline{
		\begin{minipage}{0.7\linewidth}
	\begin{tikzpicture}[scale=0.3]
	\begin{feynman}
	\vertex (a) at (-7,0){};
	\vertex (aa) at (7,0){};
	\vertex (am) at (-5.2,-0.5);
	\vertex (aam) at (5.2,-0.5);
	\draw (a) circle (70pt);
	\draw (aa) circle (70pt);
	\filldraw (am) circle (3pt);
	\filldraw (aam) circle (3pt);
	\def\myshift#1{\raisebox{-0.5ex}}
\vertex at (a) {$\mathcal{T}_1$};
\vertex at (aa) {$\mathcal{T}_2$};
	\diagram* {
		(am)--[boson,  edge label'=$\mathcal{G}$](aam)
	};
	\end{feynman}
	\end{tikzpicture}
\end{minipage}
}
\medskip

From the point of view of observables in $\mathcal{T}_1$, like scattering amplitudes between modes within this sector, the existence of $\mathcal{T}_2$ is revealed at leading order and at energies below $m_\star$, through the modification of the self-energies of the gauge bosons that belong to $\mathcal{G}$. 

The quadratic part of the action of the gauge bosons is given by
\be
S = \int\frac{d^4p}{(2\pi)^4} \frac12 V_i^\mu \Delta_{ij}(p^2) V_{j,\mu}
\ee
where we neglect the terms proportional to $p^\mu p^\nu$, and $\Delta_{VV^\prime}(q^2)$ is given by the $\eta^{\mu\nu}$ part of the propagator,
\be
\eta^{\mu\nu}\Delta_{ij}(q^2) + \ldots = -i \int d^4x e^{iq\cdot x}\braket{0 | V_i^\mu(x) V_j^{\nu}(0)|0} 
\ee
The fields $V_{i,\mu}$ interpolate the physical gauge bosons at low energies and couple to the currents of the $\mathcal{T}_1$ theory as the gauging of the global symmetry but, as will become clear in explicit examples, they are generally not mass eigenstates. 
By expanding the two-point functions in powers of $q^2$,
\be
\Delta_{ij}(q^2) = \Delta_{ij}(0) + q^2 \Delta_{ij}^\prime (0) + \frac{q^4}{2}\Delta^{\prime\prime}_{ij}(0)
+\ldots
\label{eq:Deltaexpansion}
\ee
we can classify the leading effects of the dynamics of $\mathcal{T}_2$ at long distances.

It is useful to interpret Eq.~\ref{eq:Deltaexpansion} in terms of local operators in an EFT expansion. For instance, the local operator $\mathcal{O}_{2F} = D_\mu F_{\mu\nu}D_\rho F_{\rho\nu}$, with $F_{\mu\nu}$ being a $U(1)$ field strength, is potentially generated by $\mathcal{T}_2$ at energies $q^2\ll m_\star^2$, and its dynamics parametrized by $\mathcal{L}\supset -\frac{c_{2F}}{2m_\star^2}\mathcal{O}_{2F}$ lead to the identification of the $q^4$ term in the expansion of  Eq.~\ref{eq:Deltaexpansion} with the Wilson coefficient $c_{2F}$,  $\Delta^{\prime\prime}_{FF}(0)=c_{2F}$. In a similar way, higher derivative operators are identified with coefficients of higher powers of $q^2$.

\smallskip

We can write the gauge boson self-energy in terms of the dynamics of $\mathcal{T}_2$. Denoting by $J^\mu_i$ the conserved current of the sector $\mathcal{T}_2$ associated with the vector boson $V_{i,\mu}$, the two point function of the vectors is proportional to the two-point correlator of the global symmetry current
\bea\nn
\Pi^{\mu\nu}_{ij}(q) &=& i\int d^4x\,e^{iq\cdot x} \braket{0| T J^\mu_i(x) J^\nu_j(0) |0}\\
&=&
(q^2\eta^{\mu\nu}-q^\mu q^\nu)\Pi_{ij}^T(q^2) - q^\mu q^\nu \Pi_{ij}^L(q^2) ~~.
\label{eq:Pimunudecomposition}
\eea
where $i,j=1,\ldots,N$ runs over the number of gauge bosons in $\mathcal{G}$, and we have decomposed the vacuum polarization in a transverse and a longitudinal component. 
We are interested in the ungauged limit $g\to 0$ of the correlator, 
so that the vacuum polarization is computed at leading order in $g$ but to all orders in the dynamics of $\mathcal{T}_2$.
In this limit, 
the current-current correlator affords a Källen-Lehmann \cite{Kallen:1952zz,Lehmann:1954xi} spectral representation. This is obtained by inserting a complete set of states
\bea\nn
& \braket{0| J_i^\mu(x) J_j^\nu (0) |0}
 = \\\nn
&   \int_0^\infty d\mu^2 \,\int\frac{d^4p}{(2\pi)^4}(2\pi)\theta(p^0)\delta(p^2-\mu^2) e^{-i p\cdot x} \, \rho_{ij}^{\mu\nu}(p)\\
\label{eq:jjKL}
\eea
where we have defined the spectral density $\rho^{\mu\nu}_{ij}(p)$ to be given by
\bea\nn
&(2\pi) \theta(p^0) \rho_{ij}^{\mu\nu}(p) = \sum_\alpha \delta^{(4)}(p-p_\alpha) \braket{0|J_i^\mu|\alpha}\braket{\alpha|J_j^\nu|0}\\\nn
&=
(2\pi) \theta(p^0) \left[  (p^\mu p^\nu-p^2\eta^{\mu\nu})\rho_{ij}^T(p^2)+ p^\mu p^\nu \rho_{ij}^L(p^2)  \right].
\eea
Lorentz invariance allows to decompose the spectral density into the longitudinal and transverse components. 
By evaluating the time and spatial components of the spectral density $\rho^{00}_{ij}$ and $\rho^{xx}_{ij}$ in the frame where $p^\mu = (\sqrt{p^2},\vec{0})$, one verifies that both longitudinal and transverse spectral densities are positive definite matrices, 
\be
\rho_T(p^2)\succ 0\,,\quad\quad \rho_L(p^2)\succ 0 ~~.
\ee
This is equivalent to the statement that any linear combination of currents $\sum_i\alpha_iJ^\mu_i$ has a positive spectral density associated to its two-point correlator for any choice of $\vec{\alpha}$.

After taking the time-ordered product, the advanced propagator in Eq.~\ref{eq:jjKL} changes to the Feynman propagator, which after taking the Fourier transform leads to
\bea\nn
\Pi_{ij}^{\mu\nu}(q) &=&
\int_0^\infty d\mu^2 \frac{q^\mu q^\nu - \mu^2 \eta^{\mu\nu}}{q^2-\mu^2+i\epsilon}\rho_{ij}^T(\mu^2) \\
&&+ \int_0^\infty d\mu^2 \frac{q^\mu q^\nu }{q^2-\mu^2+i\epsilon}\rho_{ij}^L(\mu^2) ~~,
\eea
to be compared with Eq.~\ref{eq:Pimunudecomposition}.
From this equation we can identify the spectral densities with the imaginary part of the vacuum polarizations,
\be
\rho^{T,L}_{ij}(q^2) \,=\, \frac{1}{\pi}\text{Im}\Pi^{T,L}_{ij}(q^2) ~~.
\ee
It is possible to examine the asymptotics of the vacuum polarization by understanding the OPE limit of the current correlator \cite{PhysRev.179.1499,Bernard:1975cd}. As long as the currents $J^\mu$ are conserved currents, the OPE is dominated by the perturbative contribution $\braket{J^\mu(x)J^\nu(0)}\sim C^{\mu\nu}(x)\frac{1}{x^6}$, which may be understood through the RG-invariance of charge conservation, while contributions from potential condensates are less singular in $x$.\footnote{We point out that the consideration of global current-current correlators in establishing positivity was independently suggested to M. McCullough by R. Rattazzi.} This scaling implies that $\text{Im}\Pi^{T,L}(q^2)$ scales as a constant at large and positive $q^2$, which implies a logarithmic divergence of the vacuum polarization.
This reflects the fact that the vector wavefunction must be renormalized. To do so, we subtract the vacuum polarization at zero momentum and define a renormalized vacuum polarization given by
\be
\Pi^\prime(q^2)  = \Pi(q^2) - \Pi(0)\,=\, \frac{q^2}{\pi}\int_0^\infty \frac{ds}{s}\frac{\text{Im}\Pi(s)}{s-q^2} ~~,
\label{eq:renormalizedvacuumpol}
\ee
which is calculable and satisfies positivity, as advertised.

Equivalently, we could have made use of the analyticity of the vacuum polarization in the complex plane except on the positive real axis, where it develops a discontinuity due to the creation of physical states that belong to $\mathcal{T}_2$. We can define a once-subtracted vacuum polarization as
\be
\Pi^\prime(q^2)\,=\, \frac{q^2}{2\pi i} \oint \frac{dz}{z}\frac{\Pi(z)}{z-q^2}\,=\, 
\frac{q^2}{\pi}\int_{m_\star^2}^\infty \frac{ds}{s}\frac{\text{Im}\Pi(s)}{s-q^2}
\label{eq:vacuumpoldispersive}
\ee
where the contour integral encloses the poles at $z=0$ and $z=q^2$ and the second inequality comes from the discontinuity around the real axis and, crucially, neglecting the contribution from the $z\to\infty$ region. 

By expanding $\Pi^\prime(q^2)$ in powers of $q^2$, $\Pi^\prime(q^2)=\sum_{n=1} q^{2n}\Pi^{(n+1)}$, with $\Pi^{(n+1)}$ being the $n$-th derivative of $\Pi^\prime(q^2)$ at $q^2=0$, we can identify the terms in the expansion in Eq.~\ref{eq:Deltaexpansion} with the vacuum polarization, $\Delta^{(n)}=\Pi^{(n)}$. Moreover, similar to Eq.~\ref{eq:vacuumpoldispersive}, one has the dispersive representation 
\be
\Pi^{(n)}(0)\,=\, \frac{1}{2\pi i} \oint \frac{dz}{z}\frac{\Pi(z)}{z^n}\,=\, 
\frac{1}{\pi}\int_{m_\star^2}^\infty \frac{ds}{s}\frac{\text{Im}\Pi(s)}{s^n} ~~,
\label{eq:momvacuumpoldispersive}
\ee
identifying $\Pi^{(n)}$ with the $n$-th moment of a positive definite matrix distribution.
The fact that the measure is a positive definite matrix implies an infinite number of nonlinear constraints on $\Pi^{(n)}$. Such constraints are given by extending the Hausdorff moment problem, which applies to positive measures in a compact domain, to consider measures that form a positive definite matrix in a compact domain. Following similar arguments as in \cite{Bellazzini:2020cot}, see also \cite{ChoqueRivero2006}, the 
necessary and sufficient conditions for identifying the sequence of matrices $\{ \Pi^{(1)},\Pi^{(2)},\ldots,\Pi^{(n)}\}$ as moments of a positive definite measure are given by
\be
H^1 \succ 0 \,,\quad
H^2 \succ 0\,,\quad
H^1-H^2 \succ 0\,,\quad
H^2-H^3 \succ 0 ~~,
\label{eq:Hankelconstraints}
\ee
where $H^k$ is the Hankel matrix of moments, $(H^k)_{ij}=\Pi^{(i+j+k-1)}$. This has exactly the same form as the constraints emanating from the Hausdorff moment problem, with the important remark that the elements $\Pi^{(i)}$ of the Hankel matrix are themselves matrices. Eq.~\ref{eq:Hankelconstraints} represents, therefore, the optimal constraints on a given set of moments of a matrix of vacuum polarizations.

\section{Overture: The Standard Model}
\label{sec:SM}
The low energy phenomenology of QED within the SM provides illustrative instances of universal theories featuring positivity of the \textit{oblique} corrections. We consider the SM below the QCD and EW scales and discuss the resulting UV impact on the photon self-energy. The first two examples are universal theories hence the leading effect is encoded in the dimension six operator $D_\mu F_{\mu\nu} D_\rho F_{\rho\nu}$. This operator alone, however, is unphysical unless some light matter fields are present in the theory. Otherwise, it may be removed via the equations of motion or, equivalently, the theory is free.\footnote{At higher derivatives, one will encounter a non-vanishing Euler-Heisenberg operator $F^4$, which will lead to nontrivial scattering. However, even in this case the $(DF)^2$ operator can be reabsorbed in a redefinition of the dimension 12 operator $F^6$.} This makes manifest the crucial role played by the matter to which the photon couples and hence the positivity of the self-energy will depend not only on how the QED might be embedded in a larger group in the UV, but also on how the matter the photon couples to is embedded within the UV group.

\textit{\textbf{Beyond QED : QCD.}}---
Consider the case where the IR theory is QED at low energies, describing only electrons and photons. A prime example of a universal UV-completion is given by QCD itself. Indeed, QCD is an instance of a universal strongly coupled sector, which manifests in the scattering of electrons only through a modification of the photon's self-energy at leading order.

\medskip
\centerline{
	\begin{minipage}{0.7\linewidth}
		\centering
		\begin{tikzpicture}[scale=0.3]
		\begin{feynman}
		\vertex (a) at (-7,0){};
		\vertex (aa) at (7,0){};
		\vertex (am) at (-5.2,-0);
		\vertex (aam) at (5.2,-0);
		\draw (a) circle (70pt);
		\draw (aa) circle (70pt);
\filldraw (am) circle (3pt);
\filldraw (aam) circle (3pt);
\def\myshift#1{\raisebox{-0.5ex}}
\draw [decoration={text along path,
	text={|\sffamily\myshift|leptons},text align={center}},decorate] (-9,-.5) to [bend right=25]  (-5,-.5);
\draw [decoration={text along path,
	text={|\sffamily\myshift|QCD},text align={center}},decorate] (5,-.5) to [bend right=25]  (9,-.5);
		\diagram* {
			(am)--[boson,  edge label'=$U(1)_{em}$](aam)
		};
		\end{feynman}
	\end{tikzpicture}
\end{minipage}
}
\medskip

Positivity of the operator $D_\mu F_{\mu\nu} D_\rho F_{\rho\nu}$ stems from the textbook discussion of the so-called R-ratio \cite{Peskin:1995ev,Jegerlehner:2017gek}, $R_\text{had}=\sigma(e^+e^-\to\text{hadrons})/\sigma(e^+e^-\to\mu^+\mu^-)$.
In the language of the previous section, the gauged symmetry $\mathcal{G}$ is a $U(1)$, with $\mathcal{T}_1$ consisting of free leptons and $\mathcal{T}_2$ the QCD sector. Therefore, the vacuum polarization in Eq.~\ref{eq:Pimunudecomposition} is a single function $\Pi^{\mu\nu}(q)$ instead of a matrix. The conservation of the current implies that $\text{Im}\Pi^T(s)\to \text{const}$ for $s\to \infty$, signaling the need to renormalize the electric charge as in Eq.~\ref{eq:renormalizedvacuumpol}. The vacuum polarization $\Pi^\prime(q^2)$ is finite, and the threshold starts at the pion mass $m_\pi^2$. By considering the scattering among different flavors of $\mathcal{T}_1$, for instance $\mathcal{A}(e^+e^-\to \mu^+\mu^-)$, one finds that at leading order in the $U(1)$ coupling $e$ and to all orders in $\alpha_s$, the imaginary part of the amplitude is, on one hand, proportional to the imaginary part of the vacuum polarization $\text{Im}\Pi^T(s)$, and on the other, proportional to the total cross section of producing hadrons, i.e.\ the $\mathcal{T}_2$ states. Explicitly,
\be
e^2\,\text{Im}\Pi^T(s) = s\,\sigma(e^+e^-\to\text{hadrons}) ~~.
\ee
At large $s$, $\text{Im}\Pi^T(s)$ is calculable in perturbation theory, while the leading effects near threshold $m_\pi^2$ are described by the photon-$\rho$ mixing \cite{Kroll:1967it,OConnell:1995nse}.

\textit{\textbf{Beyond QED : EW, universal case.}}---
Now consider the case where the universal UV-completion is given by the massive gauge bosons of the full EW group.  We focus on the hypothetical possibility that the IR-matter consists only of right handed fermions, so from the UV perspective fermions have only hypercharge and are singlets of $SU(2)$. In this case, $\mathcal{T}_1$ contains free right-handed fermions and $\mathcal{T}_2$ is the EW group with gauged $SU(2)$ and a mass gap generated by the Higgs. In this case, one has a universal theory since UV effects can be encoded entirely in the photon's self-energy. Due to the universality, coefficients of local operators controlling the corrections of the photon self-energy will be positive-definite.

\medskip
\centerline{
	\begin{minipage}{0.7\linewidth}
	\centering
	\begin{tikzpicture}[scale=0.3]
	\begin{feynman}
	\vertex (a) at (-7,0){};
	\vertex (aa) at (7,0){};
	\vertex (am) at (-5.2,-0);
	\vertex (aam) at (5.2,-0);
	\draw (a) circle (70pt);
	\draw (aa) circle (70pt);
	\filldraw (am) circle (3pt);
	\filldraw (aam) circle (3pt);
	\def\myshift#1{\raisebox{-0.5ex}}
\vertex at (-7,-0.7) {$\psi_R$};
	\draw [decoration={text along path,
		text={|\sffamily\myshift|EW},text align={center}},decorate] (5,-.5) to [bend right=25]  (9,-.5);
	\diagram* {
		(am)--[boson,  edge label'=$U(1)_Y$](aam)
	};
	\end{feynman}
	\end{tikzpicture}
\end{minipage}
}
\medskip

That the theory is universal is readily observed by considering the interactions between the photon and the $Z$ boson to the right-handed currents as
\be
\mathcal{L}\,\supset\, (Y_{e_R}J_{e_R}^\mu+Y_{q_R}J_{\mu_R}^\mu) \left( g^\prime c_W A_\mu \,-\, g^\prime s_W Z_\mu  \right),
\ee
where we denote by $Y_{e_R}$ and $Y_{q_R}$ the hypercharges of the matter fields, which coincide with the electromagnetic charges. The QED coupling is given by $e\equiv g^\prime c_W$, and $g^\prime$ and $c_W$ are the hypercharge coupling and the cosine of the Weinberg angle. Written in this way, it is clear that performing the field redefinition 
\be
\label{eq:AInt}
A_\mu \,\to\, \bar{A}_\mu + t_W Z_\mu
\ee
allows us to decouple the $Z$ boson from the light fermions.
Notice that in the IR the field $\bar{A}_\mu$ still interpolates single photon states, hence we refer to it as the photon field. In the EFT, the field redefinition is equivalent to use the equations of motion.

Since we have decoupled the $Z$ from the light matter, it is clear that at leading order we do not generate four-fermion operators. Instead,
this field redefinition induces a kinetic mixing between the interpolating field and the $Z$, given by $\mathcal{L}\supset t_W D_\mu F_{\mu\nu} Z_\nu$. When integrating out the $Z$ boson, this mixing induces a modification of the photon self-energy given by
\be
\mathcal{L} \,\supset\,  -\frac{1}{2m_Z^2}\frac{e^2}{g^2c_W^2}D_\mu F^{\mu\nu} D^\rho F_{\rho\nu}.
\label{eq:photonOPRHcurrent}
\ee
Positivity of the Wilson coefficient can be understood noticing that the $Z$ boson is interpolated by the current, $\braket{0| J^\mu|Z_{\lambda}} = \frac{e}{g c_W} m_Z^2 \epsilon_{\lambda}^\mu$, which gives a delta function contribution to the spectrum at the $Z$ pole,
\be
\begin{aligned}
\begin{tikzpicture}[scale=0.3]
\begin{feynman}[every blob={/tikz/fill=gray!30,/tikz/minimum size=30pt}]
\vertex (aa) at (-3,0){};
\vertex(a) at (-1.2,0);
\vertex(b) at (1.2,0);
\vertex (bb) at (3,0){};
\vertex at (-0.75,1){$Z$};
\vertex at (3,1){$\gamma$};
\vertex at (-3,1){$\gamma$};
\diagram* {
	(aa)--[boson](a)--[boson, line width=0.5mm] (b)--[boson](bb),
};
\draw [dashed, line width=.05cm, red!] (0,1) -- (0,-1);
\end{feynman};
\end{tikzpicture} \end{aligned} \,\,\hspace{.3cm}\rightarrow \hspace{.3cm} \frac{e^2}{g^2 c_W^2} 
m_Z^2 \delta(q^2-m_Z^2)
\ee

whose normalization is positive since it is proportional to the square of the mixing.

Alternatively one can derive the same result without relying on the interpolating field. Integrating out the $Z$ boson directly generates four-fermion interactions among the hypercharged fermions $f_1$ and $f_2$, whose coefficient is proportional to $\frac{Y_{f_1}Y_{f_2}g^{\prime 2} s_W^2}{m_Z^2}$. However, given that the photon couples to matter in a way that is aligned with the $Z$, using the equations of motion all four-fermion operators can be removed in favour of the photon's self-energy in Eq.~\ref{eq:photonOPRHcurrent}.

\medskip
Going beyond tree-level is more interesting because it forces us to understand the role of negative norm states. This is because in this example, $U(1)_\text{em}$ is embedded in a larger nonabelian gauge group, and therefore one has a priori loops of ghosts to consider. 
First of all, in general, the full theory has a gauge group $\mathcal{G}_{\mathcal{T}_1\cup\mathcal{T}_2}$, which is spontaneously broken to $\mathcal{H}$. In order to get an effective action invariant under the unbroken gauge group $\mathcal{H}$ we must use a gauge fixing term that is invariant under $\mathcal{H}$ \cite{Weinberg:1980wa}, which is our particular case is $U(1)_\text{em}$. This can be done using the gauge-fixing functional
\be
f^{\hat{a}} = \frac{1}{\sqrt{\xi}} (D_\mu V^{\hat{a}}_\mu - \xi g^{\hat{a}}\sigma^{\hat{a}}_i\chi_i)
\ee
where $D_\mu$ is the $\mathcal{H}$-covariant derivative, ${}^{\hat{a}}$ are indices corresponding to the broken generators and the second term corresponds to the gauge fixing term for the goldstones $\chi_i$. The vevs $\sigma^{\hat{a}}$ are given by $\sigma^{\hat{a}}\equiv T_\Sigma^{\hat{a}}\langle
\Sigma\rangle$, with $\Sigma$ being a scalar field transforming as $\delta\Sigma=T^A_\Sigma \alpha^A \Sigma$ under $\mathcal{G}_{\mathcal{T}_1\cup\mathcal{T}_2}$ that gets a vev $\langle
\Sigma\rangle$.
This is similar to using a background field gauge for the vectors in $\mathcal{H}$. 
Going back to the EW example, the photon self-energy receives a contribution from a $W$ loop, as well as from goldstones and ghosts 

\medskip
\centerline{
	\begin{minipage}{0.32\linewidth}
		\centering
		\begin{tikzpicture}[scale=0.3]
		\begin{feynman}[every blob={/tikz/fill=gray!30,/tikz/minimum size=30pt}]
		\vertex (a) at (-3,0){};
		\vertex (aa) at (-1.2,0);
		\vertex (bb) at (1.2,0);
		\vertex (b) at (3,0){};
		\diagram* {
			(a)--[boson, edge label'=$\gamma$](aa),
			(bb)--[boson,  edge label'=$\gamma$](b),
			(aa)--[boson, half right,edge label' = $W^+$, line width=.5mm] (bb),
			(bb)--[boson ,half right, edge label' = $W^-$, line width=.5mm] (aa),
		};
		\end{feynman}
		\end{tikzpicture}
	\end{minipage}
	\begin{minipage}{0.32\linewidth}
		\centering
		\begin{tikzpicture}[scale=0.3]
		\begin{feynman}[every blob={/tikz/fill=gray!30,/tikz/minimum size=30pt}]
		\vertex (a) at (-3,0){};
		\vertex (aa) at (-1.2,0);
		\vertex (bb) at (1.2,0);
		\vertex (b) at (3,0){};
		\diagram* {
			(a)--[boson, edge label'=$\gamma$](aa),
			(bb)--[boson, edge label'=$\gamma$](b),
			(aa)--[dotted, half right,edge label' = $c$, line width=.3mm] (bb),
			(bb)--[dotted ,half right, edge label' = $c$, line width=.3mm] (aa),
		};
		\end{feynman}
		\end{tikzpicture}
	\end{minipage}
	\begin{minipage}{0.32\linewidth}
		\centering
		\begin{tikzpicture}[scale=0.3]
		\begin{feynman}[every blob={/tikz/fill=gray!30,/tikz/minimum size=30pt}]
		\vertex (a) at (-3,0){};
		\vertex (aa) at (-1.2,0);
		\vertex (bb) at (1.2,0);
		\vertex (b) at (3,0){};
		\diagram* {
			(a)--[boson,  edge label'=$\gamma$](aa),
			(bb)--[boson,  edge label'=$\gamma$](b),
			(aa)--[dashed, half right,edge label' = $\phi^+$, line width=.3mm] (bb),
			(bb)--[dashed ,half right, edge label' = $\phi^-$, line width=.3mm] (aa),
		};
		\end{feynman}
		\end{tikzpicture}
	\end{minipage}
}

We can directly compute the imaginary part of the selfenergy.  It only depends on the photon's offshell momenta $q^2$ and the $W$ mass $m_W$ through the positive dimensionless quantity $y\equiv 4m_W^2/q^2\leq 1$.  
The cut receives a contribution from the physical $W$ polarizations, but also a contribution from the unphysical polarizations, goldstone modes and ghosts.
All together, the sum of all contributions lead to a contribution to the spectral density given by
\be
\label{eq:gaugenon}
\text{Im}\Pi = 
 - \frac{1}{3}\frac{e^{2}}{64\pi} \sqrt{1-y} \, 3(7+y),
\ee
which is negative. 
From this, one obtains the vacuum polarization and therefore the coefficient of the operator $-\frac{c_{2F}}{2 \Lambda^2} D^\mu F_{\mu\nu} D_\rho F^{\rho\nu}$, which is given by the first moment of the previous (negative definite) kernel,
\be
c_{2F} \,\propto\, \frac{1}{\pi}\int_0^1 dy\frac{\sqrt{1-y}}{64 \pi}(7+y) = -\frac{1}{(4 \pi)^2} \frac{37}{30}
\ee
which corresponds to the results obtained in \cite{Henning:2014wua} and \cite{Quevillon:2018mfl}.

The negativity of this result challenges our expectations to interpret the result as being proportional to the cross section for producing a physical state.
The key point to note is that this result is invariant under the unbroken low energy gauge group $\mathcal{H}$, but is not invariant under $\mathcal{G}_{\mathcal{T}_1\cup\mathcal{T}_2}$ and is hence unphysical, incorrectly capturing the IR effects of the UV physics. Consequently, unphysical states not only contribute but dominate the cut, giving rise to a negative coefficient. This also means that the cut cannot be interpreted as a physical production process from a right-handed current.

In fact, the amplitude to produce a $W$ pair is gauge invariant, and therefore obeys the BRST relations with the consequent decoupling of unphysical states, only after the inclusion of the $Z$ exchange.  This means that only the combinations that appear in physical quantities, such as the amplitude $\mathcal{A}(e_R\bar{e}_R\to \mu_R\bar{\mu}_R)$ between two flavors of right-handed currents, are invariant and hence physical.
In this case, in the frame where the $Z$ boson does have a direct coupling to the matter fields, besides the operator which modifies the photon's self-energy one generates a four-fermion operator of the type $\bar{f}\gamma^\mu f\bar{f}\gamma_\mu f$ and operators of the type $\bar{f}\gamma^\mu f D_\nu F_{\nu\mu}$. When integrating out the EW sector, the individual coefficients are \textit{not} invariant under $\mathcal{G}$.
In our case, the four-fermion operator would be naturally identified with the diagram with two $Z$ bosons, and the current-photon operator is naturally identified with diagrams with a $Z$ and a photon coupling to each current. Since the theory is universal, the leading effect may be encoded in the single operator $D_\mu F_{\mu\nu} D_\rho F_{\rho\nu}$, whose coefficient is positive and given by the first moment of a positive distribution, both at tree-level and at one-loop.

This contribution is more easily understood in the picture where the $Z$ coupling to matter fields is removed via field redefinition. The mixing between the photon and the $Z$ must be now included in the calculation of the photon selfenergy. While the photon's direct coupling to the $W$ pair is given by $e$, the coupling via the $Z$-mixing is proportional to $t_W\,\times\,g c_W = e$, so it transpires they are of the same order and hence equally important.  

\medskip
\centerline{
	\begin{minipage}{0.32\linewidth}
		\centering
		\begin{tikzpicture}[scale=0.3]
		\begin{feynman}[every blob={/tikz/fill=gray!30,/tikz/minimum size=30pt}]
		\vertex (aaa) at (-4,0){};
		\vertex (bbb) at (4,0){};
		\vertex (a) at (-2.4,0);
		\vertex (aa) at (-1.2,0);
		\vertex (bb) at (1.2,0);
		\vertex (b) at (2.4,0);
		\diagram* {
			(aaa) -- [boson,edge label'={\footnotesize $\gamma$}](a),
			(a)--[boson,edge label'={\footnotesize $Z$}, line width=.5mm](aa),
			(bb)--[boson, edge label'={\footnotesize $Z$}, line width=.5mm](b),
			(b) -- [boson,edge label'={\footnotesize $\gamma$}](bbb),
			(aa)--[boson, half right,edge label' = {\footnotesize $W^+$}, line width=.5mm] (bb),
			(bb)--[boson ,half right, edge label' ={\footnotesize $W^-$}, line width=.5mm] (aa),
		};
		\end{feynman}
		\end{tikzpicture}
	\end{minipage}
	\begin{minipage}{0.32\linewidth}
		\centering
		\begin{tikzpicture}[scale=0.3]
		\begin{feynman}[every blob={/tikz/fill=gray!30,/tikz/minimum size=30pt}]
		\vertex (aaa) at (-4,0){};
		\vertex (bbb) at (4,0){};
		\vertex (a) at (-2.4,0);
		\vertex (aa) at (-1.2,0);
		\vertex (bb) at (1.2,0);
		\vertex (b) at (2.4,0);
		\diagram* {
			(aaa) -- [boson,edge label'={\footnotesize $\gamma$}](a),
			(a)--[boson,edge label'={\footnotesize $Z$}, line width=.5mm](aa),
			(bb)--[boson, edge label'={\footnotesize $Z$}, line width=.5mm](b),
			(b) -- [boson,edge label'={\footnotesize $\gamma$}](bbb),
			(aa)--[dashed, half right,edge label' = {\footnotesize $\phi^+$}, line width=.4mm] (bb),
			(bb)--[boson ,half right, edge label' ={\footnotesize $W^-$}, line width=.5mm] (aa),
		};
		\end{feynman}
		\end{tikzpicture}
	\end{minipage}
	\begin{minipage}{0.32\linewidth}
		\centering
		\begin{tikzpicture}[scale=0.3]
		\begin{feynman}[every blob={/tikz/fill=gray!30,/tikz/minimum size=30pt}]
		\vertex (aaa) at (-4,0){};
		\vertex (bbb) at (3.5,0){};
		\vertex (a) at (-2.4,0);
		\vertex (aa) at (-1.2,0);
		\vertex (bb) at (1.2,0);
		\diagram* {
			(aaa) -- [boson,edge label'={\footnotesize $\gamma$}](a),
			(a)--[boson,edge label'={\footnotesize $Z$}, line width=.5mm](aa),
			(bb) -- [boson,edge label'={\footnotesize $\gamma$}](bbb),
			(aa)--[dotted, half right,edge label' = {\footnotesize $c$}, line width=.4mm] (bb),
			(bb)--[dotted,half right, edge label' ={\footnotesize $c$}, line width=.4mm] (aa),
		};
		\end{feynman}
		\end{tikzpicture}
	\end{minipage}
}

By including the mixing terms in the calculation of the vacuum polarization, 
the cut is gauge invariant and the unphysical polarizations of the $WW$ loop cancel the contributions from the goldstone and ghost, leaving only the physical polarizations to contribute to the photon selfenergy, resulting in
\be
\text{Im}\Pi \,=\, \frac{e^{2}}{3} \frac{\sqrt{1-y}}{64 \pi}  \frac{(1-y)(4+20 y + 3 y^2)}{(4r-y)^2} ~~,
\ee
where we denote $r\equiv m_W^2/m_Z^2=c_W^2$
This \emph{correct} expression for $\text{Im}\Pi$ is manifestly positive and indeed equal to the production cross section of a $W^+W^-$ pair from a right handed electron current, and gives a manifestly \emph{positive} value for $c_{2F}$.

Therefore, in the case where the charged matter couples only to hypercharge, 
the interpolating field is aligned with the $U(1)_Y$ gauge boson and the $SU(2)_L$ part of the EW sector enters only through the kinetic mixing of the interpolating field with the $Z$ boson. 
The $Z$ boson exchange and mixing plays a crucial role in restoring gauge invariance and therefore unitarity of the cross section and in also providing the correct, physical, positive, Wilson coefficient for the low energy EFT.

\pagebreak

\textit{\textbf{A Transcendental Surprise}}
Explicitly calculating $c_{2F}$ one finds an expression which is rather involved, as a function of $r$.  However, at leading order in $1-r\ll 1$, one finds
\be\label{Eq:cancell}
c_{2F} \approx \frac{1}{180 \pi^2} \frac{766991-140910 \sqrt{3} \pi}{7} +\mathcal{O}(1-r)+...  ~~,
\ee
which well approximates the full expression for SM parameters. We draw the attention of the reader to the first term.  This term exhibits a peculiar transcendental cancellation at the level $3\times 10^{-4}$.
Unfortunately we have no explanation to offer for this bizarre effect. In the limit of $r\to1$ hypercharge is vanishing and so, although it is interesting that the final Wilson coefficient in this case is the first term alone, it is also the case that it is unphysical since there is no QED to speak of in that case. However, in the limit of very small hypercharge, $(1-r)\ll 1$, only the first remains. The result is compatible with what naturalness would suggest but yet resulting of this peculiar cancellation. Since the cancellation is between a rational and doubly irrational number it does not seem that symmetry could offer an explanation.

Note that we can capture the integral for $r=1$ with the general form
\bea
I_n & = & \int dy y^n \frac{\sqrt{1-y}}{(4-y)^2} \\
& = & \int dy y^n \frac{2+y}{2 (4-y)^2 \sqrt{1-y}} \left(\frac{4-y}{2+y} - 1 \right) ~~.\nonumber
\label{eq:Iintegral}
\eea
The prefactor is heavily peaked at $y\to1$, so that is the most important region of the integral.   The first term in the brackets gives the purely irrational piece and goes to $1$ as $y\to1$.  The second term gives the purely rational piece.  Across the integration region the difference of the two terms is $\mathcal{O}(1)$, but in the region with the greatest weight they cancel very efficiently, hence the final result has to be much smaller than either of the two terms on their own.

Note that this allows one to understand the cancellation between rational and irrational contributions to the integral.  However, it offers no physical explanation and so is not satisfying.

\textit{\textbf{Beyond QED : EW, nonuniversal case.}}---
A different situation occurs when considering instead a left-handed current coupled to electromagnetism. 
Considering left handed leptons, the interactions are given by
\be
\mathcal{L} \,\supset\, e J^\mu_e A_\mu \,+\,\frac{g}{2 c_W}(J^\mu_e-J^\mu_\nu) Z_\mu\,+\, \frac{g}{\sqrt{2}} J^\mu_{e\nu} W^+_\mu + h.c.,
\ee
where $J^\mu_e$, $J^\mu_\nu$ and $J^\mu_{e\nu}$ are the electron neutral current, the neutrino neutral current and the leptonic charged current, respectively. 
This is clearly a non-universal theory; no field redefinition can remove both electron and neutrino couplings to the heavy sector
and there is no identification of two theories $\mathcal{T}_1$ and $\mathcal{T}_2$ connected via weak gauging.
Despite also being of interest, since this work is concerned with universal theories we do not study this scenario further.

\section{Beyond the Standard Model}
\label{sec:BSM}

The microscopic dynamics behind EWSB must necessarily induce deviations in the SM dynamics of the EW sector. 
As advertised, a particularly interesting subset of microscopic theories are \textit{universal} 
theories, i.e.\ those theories where the \emph{leading} deviations from the SM arise from deviations in the boson self-energies (see e.g.\ \cite{Wells:2015uba}). In this context, at low energies, BSM effects in the scattering of light SM matter fields only appear as deviations in the mediating SM gauge boson propagator. 

In such theories we can identify $\mathcal{T}_1$ with the SM matter in terms of quarks and leptons, and $\mathcal{T}_2$ with an external sector
which may or may not be strongly coupled; we will give examples of both paralleling the QCD and EW examples of QED extensions in Sec.~\ref{sec:SM}. 

\medskip
\centerline{
	\begin{minipage}{0.7\linewidth}
		\centering
		\begin{tikzpicture}[scale=0.3]
		\begin{feynman}
		\vertex (a) at (-7,0){};
		\vertex (aa) at (7,0){};
		\vertex (am) at (-5.2,-0);
		\vertex (aam) at (5.2,-0);
		\draw (a) circle (70pt);
		\draw (aa) circle (70pt);
		\filldraw (am) circle (3pt);
		\filldraw (aam) circle (3pt);
		\def\myshift#1{\raisebox{-0.5ex}}
		\vertex at (-7,-0.6) {$\psi_{SM}$};
		\draw [decoration={text along path,
			text={|\sffamily\myshift|BSM},text align={center}},decorate] (5,-.5) to [bend right=25]  (9,-.5);
		\diagram* {
			(am)--[boson,  edge label'=$\mathcal{G}_{SM}$](aam)
		};
		\end{feynman}
		\end{tikzpicture}
	\end{minipage}
}
\medskip

The BSM sector modifies the two-point function of the vectors $V_i^\mu$ and $V_j^\mu$ that belong to the SM gauge group $\mathcal{G}$.
For instance, focusing on the EW sector, $V_i V_j=\{ W^\pm W^\mp , W^3W^3, W^3B, BB \}$.  This matrix is block diagonal since $U(1)_{em}$ is unbroken. Moreover, one can also consider the Higgs self-energy, as in \cite{Englert:2019zmt}. Considering the cases where QCD is also part of $\mathcal{G}$, then the gluon self-energy can also be affected by the BSM sector. 

There are 7 parameters which encode universal new physics up to order $q^4$ in the vacuum polarization of the SM gauge bosons \cite{Barbieri:2004qk}.  In addition, we have neglected the $V$ parameter since, as we will argue, it is constrained to be subleading by unitarity, plus an extra parameter controlling the Higgs self-energy \cite{Englert:2019zmt}.

\subsection{Positivity of $W$ and $Y$}

The constraints on the moments of the vacuum polarization in Eq.~\ref{eq:Hankelconstraints} lead to constraints on the space of oblique parameters.
First and foremost, Eq.~\ref{eq:Hankelconstraints} implies the positivity of the diagonal $q^4$ coefficients of the vector two-point function. Encoding the long distance effects in the local Lagrangian
\bea\nn
\mathcal{L} &=& -\frac{W}{2m_W^2}D_\mu W^a_{\mu\nu}D_\rho W^a_{\rho\nu}
\\
&-&
\frac{Y}{2m_W^2}D_\mu B_{\mu\nu}D_\rho B_{\rho\nu} ~~,
\label{eq:WYlagrangian}
\eea
we can identify $W$ and $Y$ with the EW parameters. In term of UV data, they are also given by the moment of the spectral density associated with the corresponding current,
\be
W = \frac{g^2 m_W^2}{2} \Pi_{33}^\prime(0)
\,=\,
\frac{g^2 m_W^2}{2\pi} \int_{m_\star^2}^\infty \frac{ds}{s}\frac{\text{Im}\Pi_{T,33}(s)}{s} ~~,
\label{eq:Wdispersiverelation}
\ee
and a similar expression for $Y$. Given the positivity of the spectral density, one has
\be
W\,>\,0\,,\hspace{1.5cm}
Y\,>\,0\,.
\label{eq:WYdominance}
\ee
Therefore, in universal theories consisting of a strong sector that couples to the SM by weakly gauging its global symmetries, one has positive $W$ and $Y$ parameters up to electroweak-size loops. 
Moreover, positivity and convergence of the spectral density imply that $W$ and $Y$ are the leading corrections in the energy expansion of the the vacuum polarization for all energies below the cutoff.
For instance, the nonhomogeneous constraints in Eq.~\ref{eq:Hankelconstraints} indicate that the $q^6$ terms $\Delta^{\prime\prime\prime}_{ii}$ are necessarily subleading corrections, $\Pi_{33}^\prime-m_\star^2 \Pi_{33}^{\prime\prime}\geq 0$, with the equality saturated only by a single delta function at threshold, i.e. a single
state mixing with the SM vectors. 
Therefore, for any process at some energy $E<m_\star$ well below the cutoff, the $W$ and $Y$ parameters dominate the leading effects due to the BSM states.

\medskip

The most delicate case is for minimally coupled vectors, i.e.\ whenever the SM gauge group is embedded in a larger group.  For concreteness one can consider the UV gauge group to be $SU(2)_1\times SU(2)_2$ broken down to the EW $SU(2)_L$, but the reasoning can be extended to more general cases. 

We assume the SM fermions and Higgs to be coupled to $SU(2)_2$. The group is broken to the EW one by a scalar in the bifundamental with vev $w$. The linear combination $Q^a=c_\theta A_1^a - s_\theta A^a_2$ obtains a mass $m_\star^2=\frac{w^2}{4}(g_1^2+g_2^2)\equiv \frac{w^2}{4} g_\star^2$, with $c_\theta=g_1/g_\star$ and $s_\theta=g_2/g_\star$.  The orthogonal combination $W^a=s_\theta A_1^a + c_\theta A_2^a$ remains massless and is to be identified with the SM $SU(2)_L$ group. The EW coupling $g$ is given by $g = g_1 g_2/g_\star$, and is the interaction strength among light fermion fields and the vectors $W^a$, while the coupling with the heavy states $Q^a$ is given by $g_2 s_\theta = g_2^2/g_\star$. 

The calculation of the EW gauge boson self-energies and the $W$ parameter goes trough in analogy to Sec.~\ref{sec:SM}. The $SU(2)_1\times SU(2)_2$ extension of the SM is a universal theory, in the sense that the direct interactions of the heavy modes with the light matter fields can be removed via the field redefinition
\be
W^a_\mu \,\to\, \bar{W}^a_\mu = W^a_\mu \,-\, \frac{g_2}{g_1} Q^a_\mu.
\ee
This induces a kinetic mixing between the EW bosons and their heavy partners given by
\be
\mathcal{L} \,\supset\, \frac{g_2}{g_1} Q^a_\mu D_\nu W^a_{\mu\nu}.
\ee
The interpolating field $\bar{W}^a_\mu$ is aligned with the direction of the $SU(2)_2$ gauge bosons, which is the direction in group space the light fields couple to, in analogy to Eq.~\ref{eq:AInt} where the interpolating field $\bar{A}_\mu$ is aligned with the hypercharge.

The self-energy of such an interpolating field receives a loop contribution from the heavy vectors, which have a nonabelian $QQW$ coupling $\sim g_1 c_\theta^2 s_\theta + g_2 c_\theta s_\theta^2= g$. This contribution alone is the equivalent of Eq.~\ref{eq:gaugenon} and it is gauge-dependent. This calculation coincides with the one reported in \cite{Henning:2014wua} and \cite{Quevillon:2018mfl}. There is however another contribution at one loop, given by the mixing between the interpolating field and the heavy vectors. While the mixing is of order $g_2/g_1$, the $QQQ$ vertex, for $g_1\gg g_2$, is given by $\sim g_1$, and therefore the contribution is of order $g$, like the direct $QQW$ vertex. Together, the imaginary part of the vacuum polarization is given by
\be
\text{Im}\Pi \,=\, \frac{g^2}{16\pi^2}\frac{1}{8}\frac{(1-y)^{3/2}(4+20y+3y^2)}{(4-y)^2} ~~,
\ee
which is manifestly positive, with $y=4m_\star^2/q^2$. The integrand is indeed given by the production of heavy vectors pair from a scattering of light left-handed fermions in the $g_2\to 0$ limit.
In this limit, the $t$-channel diagram scales as $\sim g^2 \frac{g_2^2}{g_\star^2}$, so it has an extra suppression of $\frac{g_2^2}{g_\star^2}$ with respect to the $s$-channel diagram. Therefore, at order $g^2$ in the electroweak coupling and to all orders in $g_\star$, 
the imaginary part of the amplitude is in one-to-one correspondence with
the self-energy and the oblique parameters are positive definite. In particular, 
\bea\nn
W &=& \frac{g^2}{16\pi^2} \frac{m_W^2}{m_\star^2} \frac{1}{8}\int_0^1 dy \frac{(1-y)^{3/2}(4+20y+3y^2)}{(4-y)^2}\\\nn
&=& \frac{g^2}{16\pi^2} \frac{m_W^2}{m_\star^2} \, \frac{8983-1630\sqrt{3}\pi}{120}
\simeq 
\frac{g^2}{16\pi^2} \frac{m_W^2}{m_\star^2} \, 0.039 ~~,\\
\eea
which is positive and significantly smaller than the result in \cite{Henning:2014wua}. Notice the $\sim 10^{-4}$ cancellation between the rational and transcendental terms.

The argument behind positivity follows from the weak gauging of a global current, so we remark that it is a statement at leading order in $g$ but to all orders in $g_\star$.  Thus, beyond leading order in the EW gauge coupling higher-order terms may not, necessarily, be positive.  However, they are sub-leading in magnitude also due to the perturbativity of the EW gauge couplings at the EW scale.

\medskip

We conclude that in any scenario where the SM vectors mix with a \textit{strong} sector via a marginal operator, the $W$ and $Y$ parameters are positive and necessarily dominate IR effects. \footnote{The same positivity argument applies to the $Z$ parameter, defined as a modification of the gluon two-point function, $\mathcal{L}\supset-\frac{Z}{2m_W^2}D_\mu G^A_{\mu\nu}D_\rho G^A_{\rho\nu}$, and probed in high invariant mass dijets \cite{Alioli:2017jdo}.}

\bigskip

\subsection{The $X$ parameter} \label{sec:Xparam}

The $S$ and $X$ parameters control the mixing between the diagonal of the $SU(2)$ group and hypercharge, via the vacuum polarization $\Pi_{3B}(q^2)$, which is given by the correlator $\braket{0| J_3^\mu(x) J^\nu_B(0)|0}$. 
Notice that one requires a Higgs vev insertion in order to generate such contributions, which implies that the OPE of the conserved currents scales only as $\sim \frac{v^2}{x^4}$. Therefore, the spectral density has a better asymptotic behaviour and the vacuum polarization receives finite contributions from the once-subtracted spectral density.

This is related to the fact that the $S$ parameter, given by $g^2\Delta_{3B}^\prime(0)$, receives a contribution from a dimension six operator $\mathcal{O}_{WB} = H^\dagger \tau^a H W^a_{\mu\nu}B_{\mu\nu}$ even if it is a $q^2$ term.
The $X$ parameter, given by $\frac12 gg^\prime m_W^2\Delta_{3B}^{\prime\prime}(0)$ and related to the local operator $\mathcal{O}_{X} = H^\dagger \tau^a H D_\mu W^a_{\mu\nu}D_\rho B_{\rho\nu}$, is the next moment of the spectral density. 
However, since the spectral density $\rho_{3B}(z)$ has no definite sign, it is not possible to rule out theories where the $X$ parameter dominates at low energies. 
By dimensional analysis, in SILH-like theories one expects $X$ to be suppressed by a factor of $v^2/f^2$ with respect to the $W$ and $Y$ parameters, and therefore only by accident one might generate a large $X$. 

Relying purely on unitarity, the constraints in Eq.~\ref{eq:Hankelconstraints} imply that the parameters controlling the off-diagonal correlators are constrained by the diagonal terms. In particular, positivity of the leading moment implies
\be
\begin{pmatrix}
	W & X \\ X & Y
\end{pmatrix}
\,\succ\,
0
\label{eq:WYXparameters}
\ee
So not only are the $W$ and $Y$ parameters positive, but $X$ is constrained by unitarity to be smaller than the geometric average of $W$ and $Y$,
\be
W Y - X^2\,>\,0\,~~.
\label{eq:WYXconstraints}
\ee
Phenomenological implications of this relation are explored in Sec.~\ref{sec:pheno}.

Typical composite models that obey a SILH power counting \cite{Giudice:2007fh} automatically satisfy the constraint $W Y - X^2\,>\,0$. Indeed, given the Lagrangian in Eq.~\ref{eq:WYlagrangian} together with
\be
\mathcal{L}\,\supset\, - \frac{X}{m_W^2}\frac{2}{v^2} H^\dagger \tau^a H D_\mu W^a_{\mu\nu}D_\rho B_{\rho\nu} ~~,
\label{eq:Xlagrangian}
\ee
a dimensional analysis based on the SILH counting leads to an estimation of
\be
\label{eq:WYdefine}
W=c_{2W}\frac{g^2 m_W^2}{g_\star^2m_\star^2}\,\quad
Y=c_{2B}\frac{g^{\prime 2} m_W^2}{g_\star^2m_\star^2}
\ee
for the $W$ and $Y$ parameters,
while the oblique parameter $X$ is obtained from the dimension 8 operator after the Higgs gets a vev,
\be
X= c_{HWB} \frac{gg^\prime }{m_\star^4}m_W^2 v^2 \,\simeq \, c_{HWB} \frac{gg^\prime }{g_\star^2}\frac{m_W^2}{m_\star^2}\frac{v^2}{f^2} ~~.
\ee
Notice that the dimension 8 operator that gives rise to the $X$ parameter is suppressed, with respect to $\mathcal{O}_{2V}$, by an extra factor of $1/f^2=g_\star^2/m_\star^2$ instead of just $1/m_\star^2$. This implies that the $X$ parameter is suppressed not by $m_W^2/m_\star^2$ but \textit{only} by $v^2/f^2$. 

It is expected that $c_{2W}$, $c_{2B}$ and $c_{HWB}$ are $\mathcal{O}(1)$ numbers, since all dimensionful quantities, i.e. masses and couplings, are taken into account. 
In terms of these order one parameters, the constraints in Eq.~\ref{eq:WYXconstraints} imply
\be
c_{2W}\,>\,0\,,\quad
c_{2B}\,>\,0\,,\quad
c_{2W}c_{2B} \,>\,c_{HWB}^2\left(\frac{v^2}{f^2}\right)^2\,,
\ee
The first two inequalities are the positivity constraints on $W$ and $Y$. The last inequality is automatically satisfied in theories with $c_{2W}\sim c_{2B}\sim c_{HWB}\sim 1$ due to the further $v^2/f^2$ suppression in the generation of the $X$ parameter.

One might envisage a theory controlled by a power counting where, besides an overall coupling $g_\star$ and scale $m_\star$, also involves a small parameter $\xi_\star\ll 1$, in such a way that while $c_{HWB}\sim 1$, one has $c_{2W}\sim c_{2B}\sim \xi_\star$.
In such scenario, a hierarchy between the vev $v$ and the compositeness scale $f$ is enforced by unitarity, which requires $v^2/f^2\leq\xi_\star$. 
However, dimension eight operators like $H^\dagger  H D_\mu W^a_{\mu\nu}D_\rho W^a_{\rho\nu}$ and $H^\dagger H D_\mu B_{\mu\nu}D_\rho B_{\rho\nu}$ contribute to the $W$ and $Y$ parameters. Scattering of a Higgs with vector bosons is bounded by unitarity, so it is reasonable that the bounds in Eq.~\ref{eq:WYXconstraints} are automatically satisfied and therefore such theories with unitary-protected hierarchies are not possible. 
Nonetheless, if in these theories $v^2/f^2$ is \emph{accidentally} small enough, $v^2/f^2\lesssim \xi_\star$, then the leading contributions to $W$ and $Y$ come from dimension 8 operators and provide a realization of an scenario where $WY-X^2\geq 0$ might saturate.
This is speculative, thus exploration of such models is left for future work.

\subsection{Lacking Positivity?}
So far we have assumed, besides the universality of the theory, that the SM gauge fields weakly gauge a global symmetry.  Here `weakly' is critical in the argument behind positivity, which was perturbative in nature.  Now consider how the arguments behind positivity may break down.

\textit{\textbf{Non-minimal coupling.}}
One obvious example is if the universal BSM sector is nonminimally coupled to the SM gauge bosons, hence not entirely through a weak gauging of a conserved current.  For example, consider the scenario presented in \cite{Liu:2016idz}, which consists of a strong sector with a global symmetry $\mathcal{G}_{\mathcal{T}_2}$, which includes the EW group, and a gauged $U(1)^N$ symmetry giving rise to $N$ photons with coupling $g_\star$. A subgroup $\mathcal{G}$ of $\mathcal{G}_{\mathcal{T}_2}$ is gauged, such that $\text{dim}(\mathcal{G})=N$ and the $U(1)^N$ are in the adjoint of $\mathcal{G}$. This gauging implies that $[D_\mu,D_\nu] = \epsilon_A F_{\mu\nu}$, and that light matter couples with a strength $g=\epsilon_A g_\star$, ensuring the universality of the weak gauge interactions. 
However, multipole interactions involving field strengths are controlled by a scale $m_\star$ and a strong coupling $g_\star$.
The dynamics of the model at energies below the strong sector scale $m_\star$ are given by
\be
\mathcal{L} = \frac{m_\star^4}{g_\star^2} L\left( \frac{D_\mu}{m_\star}, \frac{ F_{\mu\nu}}{m_\star^2}  \right)\,-\,\frac{1}{4g_\star^2}F^2_{\mu\nu} + i \bar{\psi} \gamma^\mu (\partial_\mu + \epsilon_A T_i A^i_\mu )\psi
\ee
with $\epsilon_A$ given by $g/g_\star$ and $g^\prime/g_\star$ for $SU(2)_L$ and $U(1)_Y$ respectively.

If the dimension of the operator $\mathcal{O}^{\mu\nu}$ in $\mathcal{L}\supset \epsilon_F F_{\mu\nu}\mathcal{O}^{\mu\nu}$ is $d_{\mathcal{O}}>2$,\footnote{This is the case, indeed, if $\mathcal{O}^{\mu\nu}$ belongs to an interacting conformal field theory.}
the OPE limit of the correlator $\braket{0| \partial_\mu \mathcal{O}^{\mu\nu}(x)\partial_\rho \mathcal{O}^{\rho\nu}(0)|0}$ is given by $\sim x^{-2d_{\mathcal{O}}-2}$.
As long as $d_{\mathcal{O}}<3$, one has that $\rho_{ii}(q^2)/q^2 \to 0$ as $q^2\to \infty$, and the dispersion relation in Eq.~\ref{eq:vacuumpoldispersive} converges and the results of the previous section hold.
However, if $d_{\mathcal{O}}\geq 3$, then the once-subtracted dispersion relation for the vacuum polarization $\Pi(q^2)$ diverges, and one needs an extra subtraction.  Equivalently, in such theories the $W$ parameter is not calculable, and is defined experimentally through measuring $\Delta^{\prime\prime}$ and employing a suitable counterterm, similarly to the gauge couplings $\Delta^{\prime}=1/g$. 

\textit{\textbf{Non-physicality of the two-point function.}}
A further potential subtlety arises when considering non-abelian currents, since in this case the two-point function is not gauge invariant and only the limit where one has a global symmetry gives a physical two-point correlation function.
A relevant historical point was the argument that QFT does not admit asymptotically free theories because the spectral representation of the two-point correlator implies the positivity of the beta function \cite{Kallen:1952zz,tHooft:1998qmr}. However, non-abelian gauge theories outflank this argument since the two-point correlator is nonphysical \cite{Gross:1973id}. In fact, the gauge-invariant beta function can be computed from the two-point function only after gauge fixing to the background field gauge \cite{Grinstein:1988wz}. 
Alternatively the beta function can be computed from a physical process, for instance from computing the potential between external sources \cite{Fischler:1977yf}, where gauge invariance requires that besides the single $s$-channel gluon exchange with a gluon loop, one receives contributions from triangle and box diagrams.  Ultimately, only the sum of contributions is physical.

\textit{\textbf{Amplitude approach.}}
Both aspects may be better understood by restricting to physical observables.  Consider the forward scattering of different flavours of weakly gauged matter. 
At leading order in the gauge coupling, scattering occurs via exchange of an $s$-channel vector boson. The charges of the initial and final state can be arranged so that the exchanged vector is any element of the EW group, such that the scattering at energies well below the cutoff of the theory, $s\ll m_\star^2$, is given by\footnote{For clarity, in this section we focus the discussion on the $W$ parameter, but the argument trivially extends to the general case.}
\be
\mathcal{A}(s)\,=\, g^2 \frac{s}{s-m_W^2} \left( \,C\,+\, W \frac{s-m_W^2}{m_W^2} \,+\, \dots\right)\, ~,
\label{eq:ampExternalCurrent1}
\ee
where the non-energy growing term $C = -1-\frac{2t_W^2-1}{1-t_W^2}W$ comes from the SM contribution and the contribution from the redefinition of the SM input parameters, \cite{LHCHiggsCrossSectionWorkingGroup:2016ypw,Farina:2016rws}, and the contact interaction from the leading EFT correction due to the local operator $\mathcal{L}\supset -\frac{W}{2m_W^2} (D_\mu W^a_{\mu\nu})^2$.

The crucial point is that even above $m_\star$, at leading order in the gauge coupling $g$, the scattering takes place exclusively in the $\ell=1$ partial wave.   Higher partial waves are generated only at higher orders in $g$.  Even if the strong sector leads to an $\mathcal{O}(1)$ modification of the photon propagator due to the two-point correlator modifying the vacuum polarization, the $s$-channel diagram is of order $g^2$.

This implies that in the $s$ complex plane, $\mathcal{A}$ has a single branch cut along $s\geq m_\star^2$, with no $u$-channel discontinuity in the forward limit. The discontinuity along the cut is given by the imaginary part of the amplitude, which is positive, and is identified with the production cross section of BSM states. Secondly, since in this limit the scattering takes place through a single partial wave, the amplitude can only scale as a constant at large $s$ and therefore has a faster-than-Froissart convergence as $\mathcal{A}(s)/s\to 0$ for $s\to\infty$.

\begin{figure}
	\centering
	\includegraphics[width=0.9\linewidth]{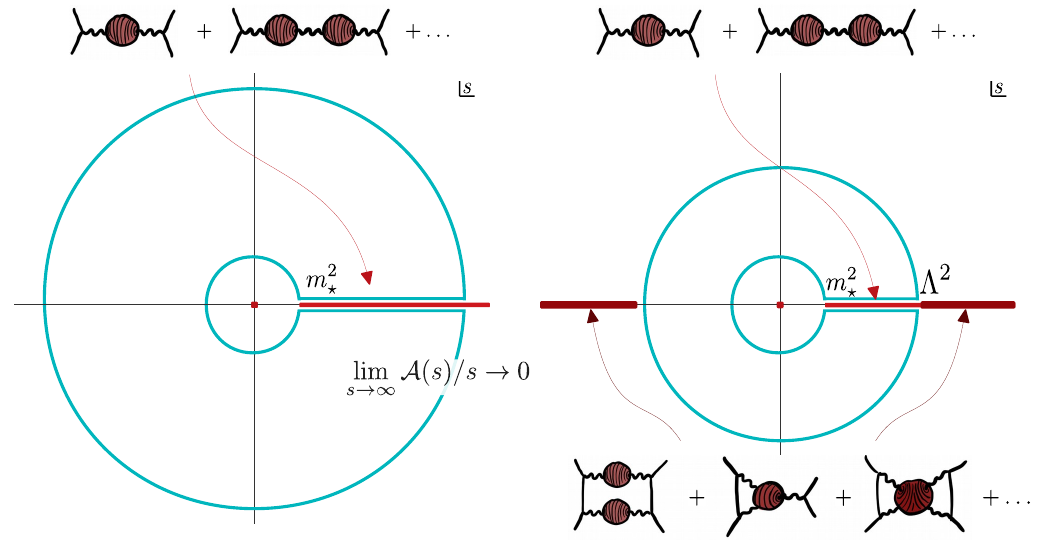}
	\caption{Left: analytic structure of the scattering amplitude between weakly gauged matter of different flavors at LO in the gauge coupling. At $m_\star^2$ the amplitude develops a branch cut due to the strong sector dynamics, but only in the $\ell=1$ partial wave. Right: the gauge coupling becomes non-perturbative above the scale $\Lambda^2$ and diagrams populating the rest of partial waves and channels are no longer negligible.}
	\label{fig:analyticstructure}
\end{figure}

The analytic structure for the amplitude is given in the left of Fig.~\ref{fig:analyticstructure}. 
The poles and branch cut due to the mixing with the strong sector start at $s=m_\star^2$. This allows to consider the quantity $\mathcal{A}(z)/z^2$ integrated along the contour in the figure, which leads to a dispersive representation for $W$ of the form
\be
W\,=\,\frac{1}{g^2}\frac{m_W^2}{2\pi i} \int_\mathcal{C} \frac{dz}{z}\frac{\mathcal{A}(z)}{z}
\,=\,
\frac{1}{g^2}\frac{m_W^2}{\pi} \int_{m_\star^2}^\infty \frac{dz}{z}\frac{\text{Im}\mathcal{A}(z)}{z}
\ee
in correspondence with Eq.~\ref{eq:Wdispersiverelation}.

\textit{\textbf{Loss of perturbativity.}}

The advantage of this perspective is that it allows to understand what happens away from the $g\to 0$ limit and for generic, non-conserved, currents. The raison d'être of both assumptions is to be able to neglect scattering through higher partial waves. We can explore therefore the scenarios that would have higher partial waves become non-negligible above some scale $\Lambda$.

The $g\to 0$ limit is still required in the IR, such that the $W$-parameter can dominate the dynamics.  However, one might imagine a situation where the strong sector induces the EW interactions to lose asymptotic freedom. At the scale $\Lambda$ at which $g(\Lambda)\sim 4\pi$, perturbativity is lost and all diagrams enter in the scattering with comparable importance.  The cut for $s>\Lambda$ contains all partial waves and thus develops a $u$-channel discontinuity as well. The analytic structure gets modified as in Fig.~\ref{fig:analyticstructure}. 
The situation in a cartoon is the following.

\medskip
\centerline{
	\begin{minipage}{0.8\linewidth}
	\centering
	\begin{tikzpicture}[scale=0.3]  
	\begin{feynman}
	\vertex (a) at (-7,0){};
	\vertex (aa) at (7,0){};
	\vertex (am) at (-5.2,-0);
	\vertex (aam) at (5.2,-0);
	\draw (a) circle (70pt);
	\draw (aa) circle (70pt);
	\filldraw (am) circle (3pt);
	\filldraw (aam) circle (3pt);
	\def\myshift#1{\raisebox{-0.5ex}}
	\vertex at (-7,-0.6) {$\psi_{SM}$};
	\draw [decoration={text along path,
		text={|\sffamily\myshift|BSM},text align={center}},decorate] (5,-.5) to [bend right=25]  (9,-.5);
	\diagram* {
		(am)--[boson,  edge label'=$\mathcal{G}_{SM}$](aam)
	};
	\vertex (a) at (-7,-6){};
	\vertex (aa) at (7,-6){};
	\vertex (am) at (-5.2,-6);
	\vertex (aam) at (5.2,-6);
	\draw (-4.8,1.1-6) arc (27.5:332.5:70pt);
	\draw (4.8,-1.1-6) arc (180+27.5:360:70pt);
	\draw (4.8,1.17-6) arc (180-27.5:0:70pt);
	\draw[ decoration={snake,amplitude=.5mm}, decorate] (-6,1.5-6) arc (270-17.2:270+17.2:20cm);
	\draw[ decoration={snake,amplitude=.5mm}, decorate] (6,-1.5-6) arc (90-17.2:90+17.2:20cm);
	\def\myshift#1{\raisebox{-0.5ex}}
	\vertex at (-7,-0.6-6) {$\psi_{SM}$};
	\vertex at (0,-1.4-6) {$\mathcal{G}_{SM}$};
	\draw [decoration={text along path,
		text={|\sffamily\myshift|BSM},text align={center}},decorate] (5,-.5-6) to [bend right=25]  (9,-.5-6);
	\draw [->] (11,1) -- (11,-7);
	\vertex at (12.5, 0) {$m_\star$};
	\vertex at (12.5,-6) {$\Lambda$};
	\end{feynman}
	\end{tikzpicture}
\end{minipage}
}
\medskip

It is possible to analyze the $W$ parameter dispersively via the contour in the right of Fig.~\ref{fig:analyticstructure}. The arc at $m_\star$ extracts the $W$ parameter. This is related to the positive discontinuity between $m_\star$ and $\Lambda$, plus the contribution from the arc at $\Lambda$. The latter is not calculable, but can be estimated via dimensional analysis. The overall coupling must be proportional to $\sim 4\pi$, since by definition it is defined as the scale where the matter fields are strongly coupled. Moreover, by dimensional analysis the arc must be proportional to $1/\Lambda^2$. The overall sign is undetermined. Therefore, we arrive at the relation 
\be
W - \eta \frac{16\pi^2}{g^2} \frac{m_W^2}{\Lambda^2}  = 
\frac{1}{g^2}\frac{m_W^2}{\pi} \int_{m_\star^2}^{\Lambda^2} \frac{dz}{z}\frac{\text{Im}\mathcal{A}(z)}{z} ~~.
\label{eq:arcatLambda}
\ee
Where Sign$(\eta)=\pm 1$ to make explicit that this second term has no definite sign.
Under the assumption that up to $\Lambda$ there is a single partial wave, NDA fixes the parameter $\eta$ to be $\eta\sim \mathcal{O}(1)$.
The right hand side of Eq.~\ref{eq:arcatLambda} is positive. Therefore, as long as $W$ dominates the left hand side of the relation, $W$ must be positive. It is useful to use dimensional analysis for $W$ as well and estimate it as $W\sim m_W^2/m_\star^2$ as in the classes of theories we just discussed. This way, positivity of the $W$ parameter is ensured as long as, parametrically, 
\be
\Lambda \,>\, m_\star \frac{4\pi}{g}\,.
\label{eq:cutoff}
\ee
In other words, under such estimate, if strong coupling is reached at arbitrary high scales, decoupling guarantees that such dynamics does not affect the $W$ parameter, which is positive. In practical terms, if Eq.~\ref{eq:cutoff} is satisfied, the large arc at $\Lambda$ can be set to zero and the conclusions from the previous section hold true.
Otherwise $\Lambda$ and the $W$ parameter has no determined sign.

There is, in fact, a large quantity that may potentially alter the $\eta\sim \mathcal{O}(1)$ estimate, which is the largest partial wave $L$ that enters in the amplitude, estimated to be $L\sim \frac{\Lambda}{m_\star}\log\frac{\Lambda^2}{m_\star^2}$ \cite{Froissart:1961ux}. 
Requiring perturbativity in $g$ above $m_\star^2$ in order to neglect the $u$-channel cut in the dispersion relation, leads to an estimate for the arc at $\Lambda$ given by $g^2(\Lambda)/\Lambda^2\frac{g^2(\Lambda)}{16\pi^2} L$, where $g^2(\Lambda)$ is the EW coupling at $\Lambda$ and the $L$ factor comes from the sum over partial waves. That both arcs become comparable requires $\Lambda\sim m_\star g^4(\Lambda)/(16\pi^2 g^2)$. By assuming $g(\Lambda)\sim 4\pi$, this estimate extends the scale $\Lambda$ by another factor $4\pi/g$ with respect Eq.~\ref{eq:cutoff}. Therefore, the fact that the electroweak coupling should run into strong coupling shortly after the $m_\star$ scale is qualitatively unmodified.
This plausibility argument can be systematically refined by considering the one-loop amplitude and dispersion relations to constrain and control the growth of the partial waves.

\section{Phenomenological discussion}\label{sec:pheno}

Positivity constraints on universal theories have a dramatic impact on the interpretation of collider data.
In this section we study the implications of positivity on current constraints and on future projections for the oblique parameters.

\subsection{Positivity of $W$ and $Y$}
The electroweak parameters $W$ and $Y$ modify the high energy behavior of the EW boson propagators, giving modifications in the high-energy tails of neutral and charged Drell-Yan processes \cite{Farina:2016rws,Alioli:2017nzr}.\footnote{Notice, also, the possible role of PDF fits for the interpretation of the Drell-Yan tails within the EFT \cite{Hammou:2023heg}.} 
As of now, the most stringent constraint on the $W$ parameter is given by the CMS collaboration analysis of lepton + MET \cite{CMS:2022krd}. Interestingly, less events than expected are measured at high invariant masses in both electron and muon channels. The collaboration interprets the data in term of the oblique parameter $W$, resulting in the constraint\footnote{CMS \cite{CMS:2022krd} normalization of the oblique parameter coincides with \cite{Farina:2016rws} and Eq.~\ref{eq:WYdefine}.}
\be
W = -1.2^{+0.5}_{-0.6}\,\times\, 10^{-4}~~.
\label{Eq:WboundNoPos}
\ee
This value is compatible with $W=0$ only at the $2\sigma$ level\footnote{The most recent analysis of the ATLAS collaboration of the same channel is found in \cite{ATLAS:2019lsy}, and also seem to report a deficit of events at large transverse mass. However, they provide no interpretation in terms of the $W$ parameter.}.

Constraints on $W$ are readily translated into constraints on the mass $m_\star$ and the coupling $g_\star$ characterizing universal theories. Assuming a power counting $W=\frac{g^2}{g_\star^2}\frac{m_W^2}{m_\star^2}$, the absolute value of the 95\% C.L. region translates to a lower bound for the scale of new physics given by $m_\star > (g/g_\star) \,5.4 \text{ TeV}$.
However, such models lead to positive $W$ parameters, and therefore this interpretation is inconsistent. 

Assuming a gaussian likelihood for the bound on $W$, we impose the prior $W>0$ corresponding to the expectation in universal theories and obtain
\be
W\,\in\, [0,0.31]\,\times\,10^{-4}\, \text{ at } 95\%\,\text{CL}\,.
\ee
The fact that the data shows a two-sigma negative excess leads to an order of magnitude stronger bounds than the ones naively expected. 
In particular, it is only with the $W>0$ prior that the interpretation $W=\frac{g^2}{g_\star^2}\frac{m_W^2}{m_\star^2}$ is meaningful. With the new bound we observe that current data puts a very strong constrain on universal resonances given by
\be
m_\star > (g/g_\star) \,14.3\text{ TeV},
\ee
which requires moderately large $g_\star$, and therefore small mixing, to lie in direct reach of LHC or future colliders.
\begin{figure}
	\begin{center}
		\includegraphics[width=0.9\linewidth]{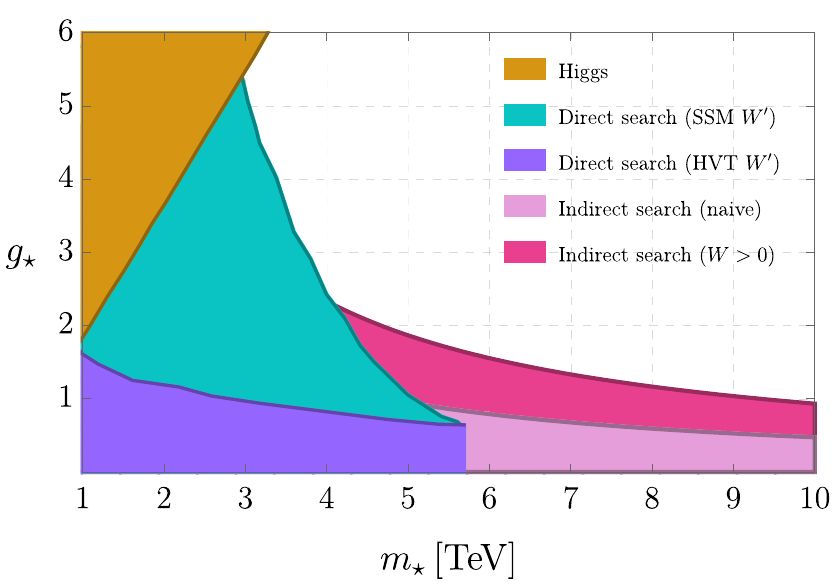}
	\end{center}
	\caption{Regions in the $g_\star - m_\star$ plane excluded at 95\%CL. In orange, the region excluded by Higgs precision measurements \cite{CMS:2018uag}. In cyan, blue and pink the regions excluded by the charged Drell-Yan data analysed in \cite{CMS:2022krd}, see their Fig~11. In red, we impose the positivity constraint $W>0$, which leads to stronger bounds, given by $W\,\in\, [0,0.31]\,\times\,10^{-4}\, \text{ at } 95\%\text{CL}$ or $m_\star > (g/g_\star) \,14.3\text{ TeV}$.}
	\label{fig:CMSplotRecast}
\end{figure}
The analysis of \cite{CMS:2022krd} interprets the constraints in the $g_\star - m_\star$ plane, as is customary in such searches. The idea is to use dimensional analysis to estimate the size of dimension six coefficients under the assumption that new dynamics is dominated by one scale, $m_\star$, and one coupling, $g_\star$ \cite{Giudice:2007fh}. 
Under such assumptions, Higgs dynamics receives modifications proportional to $\frac{g_\star^2}{g^2}\frac{m_W^2}{m_\star^2}=\frac{v^2}{f^2}$. Therefore, Higgs precision data 
\cite{CMS:2018uag}
leads to the bounds shown in Fig.~\ref{fig:CMSplotRecast}. 
In the same figure, we show the bounds from resonant searches of \cite{CMS:2022krd}, see their Fig~11.
In pink, the bounds assuming $W=\frac{g^2}{g_\star^2}\frac{m_W^2}{m_\star^2}$ and using the absolute value of the constraint in \cite{CMS:2022krd} are shown.
We remark that such universal models predict a positive $W$ parameter, and therefore this particular interpretation of the data must include such prior in the fit. After doing so, the excluded region is indicated by the red band.

\textit{\textbf{Lessons from negativity.}}
Assuming universal theories leads to $W>0$ and therefore stronger constraints. However, it is clear that there is a two-sigma tension between such hypothesis and the data, so we should explore the alternatives.

The simplest assumption to drop in order to get a deficit of events is the assumption of universality. This forces us to interpret the data in terms of individual four-fermion operators. At dimension six, the only one that modifies charged current Drell-Yan is
\be
\mathcal{L}\,\supset\, -c^{(3)}_{\ell q}\frac{g^2}{2 m_W^2} \, \bar{\ell}_L\gamma^\mu \sigma^I \ell_L \,\bar{q}_L \gamma_\mu \sigma^I q_L.
\ee
In a universal theory, the Wilson coefficient is determined by the $W$ parameter via the equations of motion,  $c^{(3)}_{\ell q} = W$.
In a non-universal theory, there is no a priori link between this coefficient and the vector boson self-energy, and therefore positivity of $c^{(3)}_{\ell q}$ is lost.

To explain the deficit of events in the high energy data consistently within the EFT, we need a \textit{tree-level size} operator. The reason for this is that the deficit of events comes from a transverse mass of $\sim 3\text{ TeV}$. Assuming conservatively that this implies a resonance above $ 3\text{ TeV}$ in order to have a consistent EFT, 
a tree-level resonance gives $c^{(3)}_{\ell q}\sim \frac{y^2}{g^2}\frac{m_W^2}{m_\star^2}$ which gives the correct size of the effect for $m_\star \sim 3\text{TeV}$ and $y\sim g$. If the leading effect were at loop level, then $c^{(3)}_{\ell q}$ would have an extra suppression of $\frac{y^2}{16\pi^2}$ and the effect would be too small to explain the deficit.

Tree-level sized and negative $c^{(3)}_{\ell q}$ can be obtained via a heavy vector $W^\prime$ which has a direct coupling with SM fermions and its coupling to either leptons or quarks has a different sign than in the SM. Also, scalar leptoquarks of the type $(3,1)_{-1/3},\, (3,3)_{-1/3}$ and vector leptoquarks in the reps $(3,1)_{2/3},\, (3,3)_{2/3}$ also may generate a negative $c^{(3)}_{\ell q}$
\cite{deBlas:2017xtg}. We should note that the deficit of events is observed in both electron and muon channels, forcing a lepton-flavour universal structure.

The other logical possibility, emphasized at the end of Sec.~\ref{sec:BSM}, is that the SM gauge coupling itself runs into strong coupling at some scale $\Lambda$ given by Eq.~\ref{eq:arcatLambda}, at most $\Lambda \sim \frac{4\pi}{g}m_\star$, which for $m_\star\sim 5\TeV$ is around $\Lambda\sim 100\TeV$. 
In such scenario, even if at $m_\star$ the theory is universal, deviations in Drell-Yan and potentially in dijet events come from strong dynamics at the $100\TeV$ scale which populates all partial waves and has no associated positivity constrain.

\begin{figure*}[t]
	\begin{center}
		\includegraphics[width=1\linewidth]{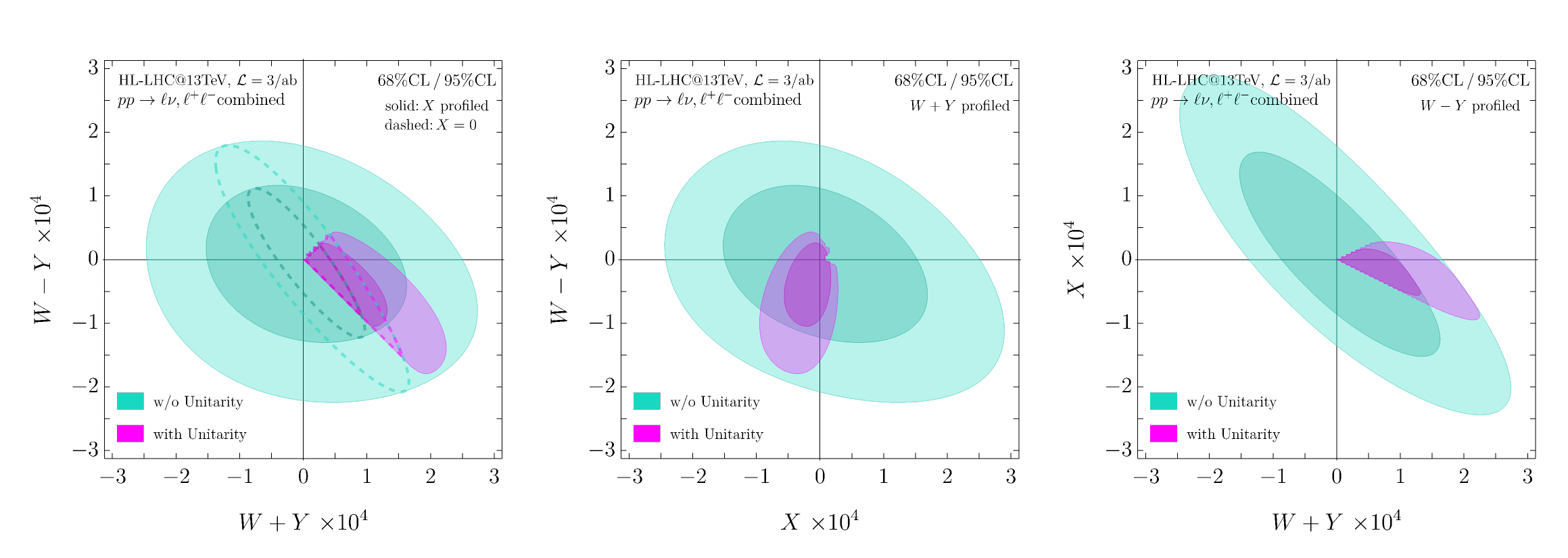}
	\end{center}
	\caption{Recast of the combination of charged and neutral Drell-Yan constraints on the individual 4-fermion interactions in \cite{Panico:2021vav} in terms of the oblique $W$, $Y$ and $X$ parameters. In cyan, the 68\% and 95\%CL regions of the naive constraints while in pink we impose the unitarity bounds of Eq.~\ref{eq:WYXconstraints}. In all plots the parameter not explicitly shown is profiled over. In the left plot, we show the constraints assuming $X=0$.}
	\label{fig:DYrecast}
\end{figure*}

\subsection{Unitarity in Drell-Yan and beyond}

We now explore how neutral and charged Drell-Yan processes explore the space of universal theories consistent with unitarity. 
The leading effects at high energies but below the EFT cutoff are parametrized by the $W$ and $Y$ parameters and also the $X$ parameter. 
As emphasized, one would expect $X$ to be parametrically smaller than $W$ and $Y$, but we will make no assumptions on the universal microscopic model, except that the theory is unitary and constraints given by the relations in Eq.~\ref{eq:WYXconstraints} hold.

In terms of local operators, $W$ and $Y$ are given in Eq.~\ref{eq:WYlagrangian}, while the operator that leads to the $X$ parameter is given in Eq.~\ref{eq:Xlagrangian}.
To make contact with the experiment, it is useful to employ the equations of motion
\bea\nn
D_\mu W^a_{\mu\nu} &=& g (H^\dagger i\overset{\leftrightarrow}{D_\nu^a} H + \sum_f \bar{f}\gamma_\nu T^a f )\,\\
D_\mu B_{\mu\nu} &=& g^\prime (Y_H H^\dagger i\overset{\leftrightarrow}{D_\nu} H + \sum_f Y_f \bar{f}\gamma_\nu f )
\eea
in order to write these operators in terms of contact four fermion operators. Only the four fermion interactions involving both quarks and leptons are relevant for the neutral and charged Drell-Yan processes. The $X$ parameter is generated by the local operator only after the Higgs takes a vev. 

There is a single relevant four-fermion operator which affects charged Drell-Yan, generated by the $W$ parameter and this is given by
\be
\mathcal{L}_{cDY}\, = \, -\frac{g^2}{2 m_W^2}\,W\, \bar{u}_L\gamma^\mu d_L \, \bar{e}_L\gamma_\mu \nu_L\,+\,h.c.\,.
\ee
Instead, neutral Drell-Yan is sensitive to several four-fermion interactions, generated by all three local gauge invariant operators. The generated interactions affecting neutral Drell-Yan are
\be
\mathcal{L}_{nDY}\, = \, -\frac{1}{m_W^2}\sum_{q,e} \mathcal{C}_{qe}\, \bar{q}\gamma^\mu q \, \bar{e}\gamma_\mu e\,,
\ee
with the sum spanning over the SM quarks and leptons fields, $q=u_L,d_L,u_R,d_R$, and $e=e_L,e_R$ as well. The coefficients $\mathcal{C}_{q e }$ are given by 
\be
\mathcal{C}_{qe}
=
g^2 T^3_{q}T^3_{e} \,W +
g^{\prime 2} Y_{q}Y_{e}\,Y +
g g^\prime (T^3_{q}Y_{e} + T^3_{e}Y_{q})\,X
\ee
as a function of $W$, $Y$ and $X$ parameters.\footnote{We use normalizations such that $T^3_{u_L}=-T^3_{d_L}=1/2$, $T^3_{u_R}=T^3_{d_R}=0$, $T^3_{e_L}=-1/2$ and $T^3_{e_R}=0$ are the isospin for each field, and $Y_{u_L}=Y_{d_L}=1/6$, $Y_{u_R}=2/3$, $Y_{d_R}=-1/3$, $Y_{e_L}=-1/2$ and $Y_{e_R}=-1$ are the hypercharges.}
Note that from the dimension six perspective, $\bar{u}_L\gamma^\mu u_L \, \bar{e}_R\gamma_\mu e_R$ and $\bar{d}_L\gamma^\mu d_L \, \bar{e}_R\gamma_\mu e_R$ necessarily have the same coefficient since they are related by $SU(2)$ invariance, but the $X$ parameter leads to a splitting among them.

We obtain the future constraints on $W$, $Y$ and $X$ recasting the HL-LHC sensitivity projections of \cite{Panico:2021vav}, for both neutral and charged Drell-Yan processes.
We interpret the constraints on the four-fermion operators in terms of the $W$, $Y$ and $X$ parameters and report the results in Fig.~\ref{fig:DYrecast}. 
In cyan, we show the 68\% and 95\%CL regions constrained by data assuming SM rates, and profiling over the parameter not explicitly shown in the plot. In pink, the same but imposing the unitarity constraints in Eq.~\ref{eq:WYXconstraints} as a prior in the fit.

In the left plot, we also show the constraints obtained by setting $X=0$, as it is commonly (and reasonably) assumed. We see that profiling over $X$ has an impact on the fit on $W$ while the fit on the $Y$ direction is more robust. However, imposing the unitarity constraints has two effects. First, it constrains the parameter space to the wedge $|W-Y|<W+Y$. Second, large $W+Y$ requires to profile over $X$ parameters outside the regime of unitarity, which also reduces the parameter space.

In the middle plot, we report the constraints on $X$ and $W-Y$. In terms of these parameters the unitary constraints are simpler in these variables, since the combination $W+Y$ sets the overall scale, while unitarity imposes
$\frac14\left(W-Y\right)^2+X^2\,<\,\frac14\left(W+Y\right)^2$,
i.e.\ that $(W-Y)/2$ and $X$ must lie in a circle of radius $W+Y$.
Therefore, in pink we show the fit including such constraints, where we profile over the \textit{radius}. 

Finally, in the right plot we show the fit in the $X$ and $W+Y$ plane. Unitarity has a dramatic impact in the fit, forcing $X$ to be in the wedge $|X|<W+Y$.

Before concluding, we should mention that the local operator that leads to the $X$ parameter generates contact interactions between four-fermions and one and two longitudinal gauge bosons, giving rise to enhanced rates in processes that look like Drell-Yan plus extra EW vectors. For instance, in the $\ell\ell VV$ final state, the signal is enhanced at both large $m_{\ell\ell}^2$ and large $s=(p_\ell+p_{\ell^\prime}+p_V+p_{V^\prime})^2$. In composite Higgs models, such signals are generated by the operator in Eq.~\ref{eq:Xlagrangian} without any suppression by $v^2/f^2$, unlike in Drell-Yan. 

It would be interesting to explore such searches since they are sensitive to a different regime of parameters of the model, dominate in theories where $v^2/f^2\ll 1$, and provide alternative test of the unitarity-driven constraints.

\section{Conclusions}
\label{sec:conc}
Basic principles of quantum mechanics and relativity impose a plethora of non-trivial constraints on possible low energy dynamics.
In this letter we presented the set of constraints on universal theories stemming from such principles. In this context, universal theories refers to a generic class of models which, in particular, include those that might describe electroweak dynamics at the microscopic level.
Thus the positivity constraints presented here have a direct impact on current and future interpretations of the data from the LHC and future colliders.

The constraints are related to the fact that deformations at low energy are encoded in the self-energy of the vector bosons that connect the two sectors of a universal theory. At leading order in the gauge coupling, the self-energy is written in terms of the current-current correlator of the strong sector. Unitarity allows to map the space of universal theories to the matrix moment problem, Eq.~\ref{eq:Hankelconstraints}, which includes and extends the constraints presented in \cite{Bellazzini:2020cot}.

The SM itself provides illustrative instances of universal theories that feature positivity of the \textit{oblique} corrections. For the relevant case of BSM models, positivity constrains the space of theories whose IR dynamics is described by the EW oblique parameters.

The argument for positivity requires the conservation of the current the gauge boson couples to, and that the gaugeless limit $g\to 0$ is a good limit. We observed that, from the perspective of scattering amplitudes, the secret role of both assumptions is to control the contribution from high partial waves. We studied the relaxation of such assumptions and concluded that positivity is maintained as long as there is a range of energies of size $4\pi/g$ during which the mixing between sectors is weakly coupled.

At the time of writing, current experimental fits on the $W$ parameter prefer negative values at the 2$\sigma$ level. This implies, when combined with the positivity constraints, bounds on the scale of new physics much stronger than expected. Alternatively, the data can be accommodated without tensions in terms of non-universal theories, since in this case there is no well-defined $W$ parameter and the definiteness of its sign is lost. The third option is that the theory is universal, but the mixing itself becomes strongly coupled at a scale not larger than $\sim 100\TeV$. Whether this is a fully realistic scenario is left for future exploration.

Future projections for the Drell-Yan data in terms of $W$, $Y$ and $X$ parameters are dramatically changed when including the unitarity constraints, as it is clear from Fig.~\ref{fig:DYrecast}.
This makes Drell-Yan a remarkable probe of unitarity in the propagation of electroweak bosons at high energy.

\subsection*{Acknowledgments}

We thank Riccardo Rattazzi for useful discussions.
The work of L.~R. is supported by NSF Grant No.~PHY-2210361 and by the Maryland Center for Fundamental Physics. 

\appendix

\bibliography{bibs} 

\begin{thebibliography}{44}%
\makeatletter
\providecommand \@ifxundefined [1]{%
 \@ifx{#1\undefined}
}%
\providecommand \@ifnum [1]{%
 \ifnum #1\expandafter \@firstoftwo
 \else \expandafter \@secondoftwo
 \fi
}%
\providecommand \@ifx [1]{%
 \ifx #1\expandafter \@firstoftwo
 \else \expandafter \@secondoftwo
 \fi
}%
\providecommand \natexlab [1]{#1}%
\providecommand \enquote  [1]{``#1''}%
\providecommand \bibnamefont  [1]{#1}%
\providecommand \bibfnamefont [1]{#1}%
\providecommand \citenamefont [1]{#1}%
\providecommand \href@noop [0]{\@secondoftwo}%
\providecommand \href [0]{\begingroup \@sanitize@url \@href}%
\providecommand \@href[1]{\@@startlink{#1}\@@href}%
\providecommand \@@href[1]{\endgroup#1\@@endlink}%
\providecommand \@sanitize@url [0]{\catcode `\\12\catcode `\$12\catcode
  `\&12\catcode `\#12\catcode `\^12\catcode `\_12\catcode `\%12\relax}%
\providecommand \@@startlink[1]{}%
\providecommand \@@endlink[0]{}%
\providecommand \url  [0]{\begingroup\@sanitize@url \@url }%
\providecommand \@url [1]{\endgroup\@href {#1}{\urlprefix }}%
\providecommand \urlprefix  [0]{URL }%
\providecommand \Eprint [0]{\href }%
\providecommand \doibase [0]{http://dx.doi.org/}%
\providecommand \selectlanguage [0]{\@gobble}%
\providecommand \bibinfo  [0]{\@secondoftwo}%
\providecommand \bibfield  [0]{\@secondoftwo}%
\providecommand \translation [1]{[#1]}%
\providecommand \BibitemOpen [0]{}%
\providecommand \bibitemStop [0]{}%
\providecommand \bibitemNoStop [0]{.\EOS\space}%
\providecommand \EOS [0]{\spacefactor3000\relax}%
\providecommand \BibitemShut  [1]{\csname bibitem#1\endcsname}%
\let\auto@bib@innerbib\@empty
\bibitem [{\citenamefont {Farina}\ \emph {et~al.}(2017)\citenamefont {Farina},
  \citenamefont {Panico}, \citenamefont {Pappadopulo}, \citenamefont
  {Ruderman}, \citenamefont {Torre},\ and\ \citenamefont
  {Wulzer}}]{Farina:2016rws}%
  \BibitemOpen
  \bibfield  {author} {\bibinfo {author} {\bibfnamefont {M.}~\bibnamefont
  {Farina}}, \bibinfo {author} {\bibfnamefont {G.}~\bibnamefont {Panico}},
  \bibinfo {author} {\bibfnamefont {D.}~\bibnamefont {Pappadopulo}}, \bibinfo
  {author} {\bibfnamefont {J.~T.}\ \bibnamefont {Ruderman}}, \bibinfo {author}
  {\bibfnamefont {R.}~\bibnamefont {Torre}}, \ and\ \bibinfo {author}
  {\bibfnamefont {A.}~\bibnamefont {Wulzer}},\ }\href {\doibase
  10.1016/j.physletb.2017.06.043} {\bibfield  {journal} {\bibinfo  {journal}
  {Phys. Lett. B}\ }\textbf {\bibinfo {volume} {772}},\ \bibinfo {pages} {210}
  (\bibinfo {year} {2017})},\ \Eprint {http://arxiv.org/abs/1609.08157}
  {arXiv:1609.08157 [hep-ph]} \BibitemShut {NoStop}%
\bibitem [{\citenamefont {Torre}\ \emph {et~al.}(2021)\citenamefont {Torre},
  \citenamefont {Ricci},\ and\ \citenamefont {Wulzer}}]{Torre:2020aiz}%
  \BibitemOpen
  \bibfield  {author} {\bibinfo {author} {\bibfnamefont {R.}~\bibnamefont
  {Torre}}, \bibinfo {author} {\bibfnamefont {L.}~\bibnamefont {Ricci}}, \ and\
  \bibinfo {author} {\bibfnamefont {A.}~\bibnamefont {Wulzer}},\ }\href
  {\doibase 10.1007/JHEP02(2021)144} {\bibfield  {journal} {\bibinfo  {journal}
  {JHEP}\ }\textbf {\bibinfo {volume} {02}},\ \bibinfo {pages} {144} (\bibinfo
  {year} {2021})},\ \Eprint {http://arxiv.org/abs/2008.12978} {arXiv:2008.12978
  [hep-ph]} \BibitemShut {NoStop}%
\bibitem [{\citenamefont {Panico}\ \emph {et~al.}(2021)\citenamefont {Panico},
  \citenamefont {Ricci},\ and\ \citenamefont {Wulzer}}]{Panico:2021vav}%
  \BibitemOpen
  \bibfield  {author} {\bibinfo {author} {\bibfnamefont {G.}~\bibnamefont
  {Panico}}, \bibinfo {author} {\bibfnamefont {L.}~\bibnamefont {Ricci}}, \
  and\ \bibinfo {author} {\bibfnamefont {A.}~\bibnamefont {Wulzer}},\ }\href
  {\doibase 10.1007/JHEP07(2021)086} {\bibfield  {journal} {\bibinfo  {journal}
  {JHEP}\ }\textbf {\bibinfo {volume} {07}},\ \bibinfo {pages} {086} (\bibinfo
  {year} {2021})},\ \Eprint {http://arxiv.org/abs/2103.10532} {arXiv:2103.10532
  [hep-ph]} \BibitemShut {NoStop}%
\bibitem [{\citenamefont {Tumasyan}\ \emph {et~al.}(2022)\citenamefont
  {Tumasyan} \emph {et~al.}}]{CMS:2022krd}%
  \BibitemOpen
  \bibfield  {author} {\bibinfo {author} {\bibfnamefont {A.}~\bibnamefont
  {Tumasyan}} \emph {et~al.} (\bibinfo {collaboration} {CMS}),\ }\href
  {\doibase 10.1007/JHEP07(2022)067} {\bibfield  {journal} {\bibinfo  {journal}
  {JHEP}\ }\textbf {\bibinfo {volume} {07}},\ \bibinfo {pages} {067} (\bibinfo
  {year} {2022})},\ \Eprint {http://arxiv.org/abs/2202.06075} {arXiv:2202.06075
  [hep-ex]} \BibitemShut {NoStop}%
\bibitem [{\citenamefont {Amoroso}\ \emph {et~al.}(2022)\citenamefont {Amoroso}
  \emph {et~al.}}]{Amoroso:2022eow}%
  \BibitemOpen
  \bibfield  {author} {\bibinfo {author} {\bibfnamefont {S.}~\bibnamefont
  {Amoroso}} \emph {et~al.},\ }\href {\doibase 10.5506/APhysPolB.53.12-A1}
  {\bibfield  {journal} {\bibinfo  {journal} {Acta Phys. Polon. B}\ }\textbf
  {\bibinfo {volume} {53}},\ \bibinfo {pages} {12} (\bibinfo {year} {2022})},\
  \Eprint {http://arxiv.org/abs/2203.13923} {arXiv:2203.13923 [hep-ph]}
  \BibitemShut {NoStop}%
\bibitem [{\citenamefont {Campbell}\ \emph {et~al.}(2022)\citenamefont
  {Campbell} \emph {et~al.}}]{Campbell:2022qmc}%
  \BibitemOpen
  \bibfield  {author} {\bibinfo {author} {\bibfnamefont {J.~M.}\ \bibnamefont
  {Campbell}} \emph {et~al.},\ }in\ \href@noop {} {\emph {\bibinfo {booktitle}
  {{Snowmass 2021}}}}\ (\bibinfo {year} {2022})\ \Eprint
  {http://arxiv.org/abs/2203.11110} {arXiv:2203.11110 [hep-ph]} \BibitemShut
  {NoStop}%
\bibitem [{\citenamefont {Barbieri}\ \emph
  {et~al.}(2004{\natexlab{a}})\citenamefont {Barbieri}, \citenamefont
  {Pomarol}, \citenamefont {Rattazzi},\ and\ \citenamefont
  {Strumia}}]{Barbieri:2004qk}%
  \BibitemOpen
  \bibfield  {author} {\bibinfo {author} {\bibfnamefont {R.}~\bibnamefont
  {Barbieri}}, \bibinfo {author} {\bibfnamefont {A.}~\bibnamefont {Pomarol}},
  \bibinfo {author} {\bibfnamefont {R.}~\bibnamefont {Rattazzi}}, \ and\
  \bibinfo {author} {\bibfnamefont {A.}~\bibnamefont {Strumia}},\ }\href
  {\doibase 10.1016/j.nuclphysb.2004.10.014} {\bibfield  {journal} {\bibinfo
  {journal} {Nucl. Phys. B}\ }\textbf {\bibinfo {volume} {703}},\ \bibinfo
  {pages} {127} (\bibinfo {year} {2004}{\natexlab{a}})},\ \Eprint
  {http://arxiv.org/abs/hep-ph/0405040} {arXiv:hep-ph/0405040} \BibitemShut
  {NoStop}%
\bibitem [{\citenamefont {Wells}\ and\ \citenamefont
  {Zhang}(2016)}]{Wells:2015uba}%
  \BibitemOpen
  \bibfield  {author} {\bibinfo {author} {\bibfnamefont {J.~D.}\ \bibnamefont
  {Wells}}\ and\ \bibinfo {author} {\bibfnamefont {Z.}~\bibnamefont {Zhang}},\
  }\href {\doibase 10.1007/JHEP01(2016)123} {\bibfield  {journal} {\bibinfo
  {journal} {JHEP}\ }\textbf {\bibinfo {volume} {01}},\ \bibinfo {pages} {123}
  (\bibinfo {year} {2016})},\ \Eprint {http://arxiv.org/abs/1510.08462}
  {arXiv:1510.08462 [hep-ph]} \BibitemShut {NoStop}%
\bibitem [{\citenamefont {Einhorn}\ \emph {et~al.}(1981)\citenamefont
  {Einhorn}, \citenamefont {Jones},\ and\ \citenamefont
  {Veltman}}]{EINHORN1981146}%
  \BibitemOpen
  \bibfield  {author} {\bibinfo {author} {\bibfnamefont {M.}~\bibnamefont
  {Einhorn}}, \bibinfo {author} {\bibfnamefont {D.}~\bibnamefont {Jones}}, \
  and\ \bibinfo {author} {\bibfnamefont {M.}~\bibnamefont {Veltman}},\ }\href
  {\doibase https://doi.org/10.1016/0550-3213(81)90292-3} {\bibfield  {journal}
  {\bibinfo  {journal} {Nuclear Physics B}\ }\textbf {\bibinfo {volume}
  {191}},\ \bibinfo {pages} {146} (\bibinfo {year} {1981})}\BibitemShut
  {NoStop}%
\bibitem [{\citenamefont {Peskin}\ and\ \citenamefont
  {Takeuchi}(1990)}]{Peskin:1990zt}%
  \BibitemOpen
  \bibfield  {author} {\bibinfo {author} {\bibfnamefont {M.~E.}\ \bibnamefont
  {Peskin}}\ and\ \bibinfo {author} {\bibfnamefont {T.}~\bibnamefont
  {Takeuchi}},\ }\href {\doibase 10.1103/PhysRevLett.65.964} {\bibfield
  {journal} {\bibinfo  {journal} {Phys. Rev. Lett.}\ }\textbf {\bibinfo
  {volume} {65}},\ \bibinfo {pages} {964} (\bibinfo {year} {1990})}\BibitemShut
  {NoStop}%
\bibitem [{\citenamefont {Sundrum}\ and\ \citenamefont
  {Hsu}(1993)}]{Sundrum:1991rf}%
  \BibitemOpen
  \bibfield  {author} {\bibinfo {author} {\bibfnamefont {R.}~\bibnamefont
  {Sundrum}}\ and\ \bibinfo {author} {\bibfnamefont {S.~D.~H.}\ \bibnamefont
  {Hsu}},\ }\href {\doibase 10.1016/0550-3213(93)90144-E} {\bibfield  {journal}
  {\bibinfo  {journal} {Nucl. Phys. B}\ }\textbf {\bibinfo {volume} {391}},\
  \bibinfo {pages} {127} (\bibinfo {year} {1993})},\ \Eprint
  {http://arxiv.org/abs/hep-ph/9206225} {arXiv:hep-ph/9206225} \BibitemShut
  {NoStop}%
\bibitem [{\citenamefont {Barbieri}\ \emph
  {et~al.}(2004{\natexlab{b}})\citenamefont {Barbieri}, \citenamefont
  {Pomarol},\ and\ \citenamefont {Rattazzi}}]{Barbieri:2003pr}%
  \BibitemOpen
  \bibfield  {author} {\bibinfo {author} {\bibfnamefont {R.}~\bibnamefont
  {Barbieri}}, \bibinfo {author} {\bibfnamefont {A.}~\bibnamefont {Pomarol}}, \
  and\ \bibinfo {author} {\bibfnamefont {R.}~\bibnamefont {Rattazzi}},\ }\href
  {\doibase 10.1016/j.physletb.2004.04.005} {\bibfield  {journal} {\bibinfo
  {journal} {Phys. Lett. B}\ }\textbf {\bibinfo {volume} {591}},\ \bibinfo
  {pages} {141} (\bibinfo {year} {2004}{\natexlab{b}})},\ \Eprint
  {http://arxiv.org/abs/hep-ph/0310285} {arXiv:hep-ph/0310285} \BibitemShut
  {NoStop}%
\bibitem [{\citenamefont {Agashe}\ \emph {et~al.}(2007)\citenamefont {Agashe},
  \citenamefont {Csaki}, \citenamefont {Grojean},\ and\ \citenamefont
  {Reece}}]{Agashe:2007mc}%
  \BibitemOpen
  \bibfield  {author} {\bibinfo {author} {\bibfnamefont {K.}~\bibnamefont
  {Agashe}}, \bibinfo {author} {\bibfnamefont {C.}~\bibnamefont {Csaki}},
  \bibinfo {author} {\bibfnamefont {C.}~\bibnamefont {Grojean}}, \ and\
  \bibinfo {author} {\bibfnamefont {M.}~\bibnamefont {Reece}},\ }\href
  {\doibase 10.1088/1126-6708/2007/12/003} {\bibfield  {journal} {\bibinfo
  {journal} {JHEP}\ }\textbf {\bibinfo {volume} {12}},\ \bibinfo {pages} {003}
  (\bibinfo {year} {2007})},\ \Eprint {http://arxiv.org/abs/0704.1821}
  {arXiv:0704.1821 [hep-ph]} \BibitemShut {NoStop}%
\bibitem [{\citenamefont {Peskin}\ and\ \citenamefont
  {Takeuchi}(1992)}]{PhysRevD.46.381}%
  \BibitemOpen
  \bibfield  {author} {\bibinfo {author} {\bibfnamefont {M.~E.}\ \bibnamefont
  {Peskin}}\ and\ \bibinfo {author} {\bibfnamefont {T.}~\bibnamefont
  {Takeuchi}},\ }\href {\doibase 10.1103/PhysRevD.46.381} {\bibfield  {journal}
  {\bibinfo  {journal} {Phys. Rev. D}\ }\textbf {\bibinfo {volume} {46}},\
  \bibinfo {pages} {381} (\bibinfo {year} {1992})}\BibitemShut {NoStop}%
\bibitem [{\citenamefont {He}\ \emph {et~al.}(2001)\citenamefont {He},
  \citenamefont {Polonsky},\ and\ \citenamefont {Su}}]{He:2001tp}%
  \BibitemOpen
  \bibfield  {author} {\bibinfo {author} {\bibfnamefont {H.-J.}\ \bibnamefont
  {He}}, \bibinfo {author} {\bibfnamefont {N.}~\bibnamefont {Polonsky}}, \ and\
  \bibinfo {author} {\bibfnamefont {S.-f.}\ \bibnamefont {Su}},\ }\href
  {\doibase 10.1103/PhysRevD.64.053004} {\bibfield  {journal} {\bibinfo
  {journal} {Phys. Rev. D}\ }\textbf {\bibinfo {volume} {64}},\ \bibinfo
  {pages} {053004} (\bibinfo {year} {2001})},\ \Eprint
  {http://arxiv.org/abs/hep-ph/0102144} {arXiv:hep-ph/0102144} \BibitemShut
  {NoStop}%
\bibitem [{\citenamefont {Cacciapaglia}\ \emph {et~al.}(2006)\citenamefont
  {Cacciapaglia}, \citenamefont {Csaki}, \citenamefont {Marandella},\ and\
  \citenamefont {Strumia}}]{Cacciapaglia:2006pk}%
  \BibitemOpen
  \bibfield  {author} {\bibinfo {author} {\bibfnamefont {G.}~\bibnamefont
  {Cacciapaglia}}, \bibinfo {author} {\bibfnamefont {C.}~\bibnamefont {Csaki}},
  \bibinfo {author} {\bibfnamefont {G.}~\bibnamefont {Marandella}}, \ and\
  \bibinfo {author} {\bibfnamefont {A.}~\bibnamefont {Strumia}},\ }\href
  {\doibase 10.1103/PhysRevD.74.033011} {\bibfield  {journal} {\bibinfo
  {journal} {Phys. Rev. D}\ }\textbf {\bibinfo {volume} {74}},\ \bibinfo
  {pages} {033011} (\bibinfo {year} {2006})},\ \Eprint
  {http://arxiv.org/abs/hep-ph/0604111} {arXiv:hep-ph/0604111} \BibitemShut
  {NoStop}%
\bibitem [{\citenamefont {Kallen}(1952)}]{Kallen:1952zz}%
  \BibitemOpen
  \bibfield  {author} {\bibinfo {author} {\bibfnamefont {G.}~\bibnamefont
  {Kallen}},\ }\href {\doibase 10.1007/978-3-319-00627-7_90} {\bibfield
  {journal} {\bibinfo  {journal} {Helv. Phys. Acta}\ }\textbf {\bibinfo
  {volume} {25}},\ \bibinfo {pages} {417} (\bibinfo {year} {1952})}\BibitemShut
  {NoStop}%
\bibitem [{\citenamefont {Lehmann}(1954)}]{Lehmann:1954xi}%
  \BibitemOpen
  \bibfield  {author} {\bibinfo {author} {\bibfnamefont {H.}~\bibnamefont
  {Lehmann}},\ }\href {\doibase 10.1007/BF02783624} {\bibfield  {journal}
  {\bibinfo  {journal} {Nuovo Cim.}\ }\textbf {\bibinfo {volume} {11}},\
  \bibinfo {pages} {342} (\bibinfo {year} {1954})}\BibitemShut {NoStop}%
\bibitem [{\citenamefont {Wilson}(1969)}]{PhysRev.179.1499}%
  \BibitemOpen
  \bibfield  {author} {\bibinfo {author} {\bibfnamefont {K.~G.}\ \bibnamefont
  {Wilson}},\ }\href {\doibase 10.1103/PhysRev.179.1499} {\bibfield  {journal}
  {\bibinfo  {journal} {Phys. Rev.}\ }\textbf {\bibinfo {volume} {179}},\
  \bibinfo {pages} {1499} (\bibinfo {year} {1969})}\BibitemShut {NoStop}%
\bibitem [{\citenamefont {Bernard}\ \emph {et~al.}(1975)\citenamefont
  {Bernard}, \citenamefont {Duncan}, \citenamefont {LoSecco},\ and\
  \citenamefont {Weinberg}}]{Bernard:1975cd}%
  \BibitemOpen
  \bibfield  {author} {\bibinfo {author} {\bibfnamefont {C.~W.}\ \bibnamefont
  {Bernard}}, \bibinfo {author} {\bibfnamefont {A.}~\bibnamefont {Duncan}},
  \bibinfo {author} {\bibfnamefont {J.}~\bibnamefont {LoSecco}}, \ and\
  \bibinfo {author} {\bibfnamefont {S.}~\bibnamefont {Weinberg}},\ }\href
  {\doibase 10.1103/PhysRevD.12.792} {\bibfield  {journal} {\bibinfo  {journal}
  {Phys. Rev. D}\ }\textbf {\bibinfo {volume} {12}},\ \bibinfo {pages} {792}
  (\bibinfo {year} {1975})}\BibitemShut {NoStop}%
\bibitem [{\citenamefont {Bellazzini}\ \emph {et~al.}(2021)\citenamefont
  {Bellazzini}, \citenamefont {Elias~Mir\'o}, \citenamefont {Rattazzi},
  \citenamefont {Riembau},\ and\ \citenamefont {Riva}}]{Bellazzini:2020cot}%
  \BibitemOpen
  \bibfield  {author} {\bibinfo {author} {\bibfnamefont {B.}~\bibnamefont
  {Bellazzini}}, \bibinfo {author} {\bibfnamefont {J.}~\bibnamefont
  {Elias~Mir\'o}}, \bibinfo {author} {\bibfnamefont {R.}~\bibnamefont
  {Rattazzi}}, \bibinfo {author} {\bibfnamefont {M.}~\bibnamefont {Riembau}}, \
  and\ \bibinfo {author} {\bibfnamefont {F.}~\bibnamefont {Riva}},\ }\href
  {\doibase 10.1103/PhysRevD.104.036006} {\bibfield  {journal} {\bibinfo
  {journal} {Phys. Rev. D}\ }\textbf {\bibinfo {volume} {104}},\ \bibinfo
  {pages} {036006} (\bibinfo {year} {2021})},\ \Eprint
  {http://arxiv.org/abs/2011.00037} {arXiv:2011.00037 [hep-th]} \BibitemShut
  {NoStop}%
\bibitem [{\citenamefont {Choque~Rivero}\ \emph {et~al.}(2006)\citenamefont
  {Choque~Rivero}, \citenamefont {Dyukarev}, \citenamefont {Fritzsche},\ and\
  \citenamefont {Kirstein}}]{ChoqueRivero2006}%
  \BibitemOpen
  \bibfield  {author} {\bibinfo {author} {\bibfnamefont {A.~E.}\ \bibnamefont
  {Choque~Rivero}}, \bibinfo {author} {\bibfnamefont {Y.~M.}\ \bibnamefont
  {Dyukarev}}, \bibinfo {author} {\bibfnamefont {B.}~\bibnamefont {Fritzsche}},
  \ and\ \bibinfo {author} {\bibfnamefont {B.}~\bibnamefont {Kirstein}},\
  }\enquote {\bibinfo {title} {A truncated matricial moment problem on a finite
  interval},}\ in\ \href {\doibase 10.1007/3-7643-7547-7_4} {\emph {\bibinfo
  {booktitle} {Interpolation, Schur Functions and Moment Problems}}},\ \bibinfo
  {editor} {edited by\ \bibinfo {editor} {\bibfnamefont {D.}~\bibnamefont
  {Alpay}}\ and\ \bibinfo {editor} {\bibfnamefont {I.}~\bibnamefont
  {Gohberg}}}\ (\bibinfo  {publisher} {Birkh{\"a}user Basel},\ \bibinfo
  {address} {Basel},\ \bibinfo {year} {2006})\ pp.\ \bibinfo {pages}
  {121--173}\BibitemShut {NoStop}%
\bibitem [{\citenamefont {Peskin}\ and\ \citenamefont
  {Schroeder}(1995)}]{Peskin:1995ev}%
  \BibitemOpen
  \bibfield  {author} {\bibinfo {author} {\bibfnamefont {M.~E.}\ \bibnamefont
  {Peskin}}\ and\ \bibinfo {author} {\bibfnamefont {D.~V.}\ \bibnamefont
  {Schroeder}},\ }\href@noop {} {\emph {\bibinfo {title} {{An Introduction to
  quantum field theory}}}}\ (\bibinfo  {publisher} {Addison-Wesley},\ \bibinfo
  {address} {Reading, USA},\ \bibinfo {year} {1995})\BibitemShut {NoStop}%
\bibitem [{\citenamefont {Jegerlehner}(2017)}]{Jegerlehner:2017gek}%
  \BibitemOpen
  \bibfield  {author} {\bibinfo {author} {\bibfnamefont {F.}~\bibnamefont
  {Jegerlehner}},\ }\href {\doibase 10.1007/978-3-319-63577-4} {\emph {\bibinfo
  {title} {{The Anomalous Magnetic Moment of the Muon}}}},\ Vol.\ \bibinfo
  {volume} {274}\ (\bibinfo  {publisher} {Springer},\ \bibinfo {address}
  {Cham},\ \bibinfo {year} {2017})\BibitemShut {NoStop}%
\bibitem [{\citenamefont {Kroll}\ \emph {et~al.}(1967)\citenamefont {Kroll},
  \citenamefont {Lee},\ and\ \citenamefont {Zumino}}]{Kroll:1967it}%
  \BibitemOpen
  \bibfield  {author} {\bibinfo {author} {\bibfnamefont {N.~M.}\ \bibnamefont
  {Kroll}}, \bibinfo {author} {\bibfnamefont {T.~D.}\ \bibnamefont {Lee}}, \
  and\ \bibinfo {author} {\bibfnamefont {B.}~\bibnamefont {Zumino}},\ }\href
  {\doibase 10.1103/PhysRev.157.1376} {\bibfield  {journal} {\bibinfo
  {journal} {Phys. Rev.}\ }\textbf {\bibinfo {volume} {157}},\ \bibinfo {pages}
  {1376} (\bibinfo {year} {1967})}\BibitemShut {NoStop}%
\bibitem [{\citenamefont {O'Connell}\ \emph {et~al.}(1997)\citenamefont
  {O'Connell}, \citenamefont {Pearce}, \citenamefont {Thomas},\ and\
  \citenamefont {Williams}}]{OConnell:1995nse}%
  \BibitemOpen
  \bibfield  {author} {\bibinfo {author} {\bibfnamefont {H.~B.}\ \bibnamefont
  {O'Connell}}, \bibinfo {author} {\bibfnamefont {B.~C.}\ \bibnamefont
  {Pearce}}, \bibinfo {author} {\bibfnamefont {A.~W.}\ \bibnamefont {Thomas}},
  \ and\ \bibinfo {author} {\bibfnamefont {A.~G.}\ \bibnamefont {Williams}},\
  }\href {\doibase 10.1016/S0146-6410(97)00044-6} {\bibfield  {journal}
  {\bibinfo  {journal} {Prog. Part. Nucl. Phys.}\ }\textbf {\bibinfo {volume}
  {39}},\ \bibinfo {pages} {201} (\bibinfo {year} {1997})},\ \Eprint
  {http://arxiv.org/abs/hep-ph/9501251} {arXiv:hep-ph/9501251} \BibitemShut
  {NoStop}%
\bibitem [{\citenamefont {Weinberg}(1980)}]{Weinberg:1980wa}%
  \BibitemOpen
  \bibfield  {author} {\bibinfo {author} {\bibfnamefont {S.}~\bibnamefont
  {Weinberg}},\ }\href {\doibase 10.1016/0370-2693(80)90660-7} {\bibfield
  {journal} {\bibinfo  {journal} {Phys. Lett. B}\ }\textbf {\bibinfo {volume}
  {91}},\ \bibinfo {pages} {51} (\bibinfo {year} {1980})}\BibitemShut {NoStop}%
\bibitem [{\citenamefont {Henning}\ \emph {et~al.}(2016)\citenamefont
  {Henning}, \citenamefont {Lu},\ and\ \citenamefont
  {Murayama}}]{Henning:2014wua}%
  \BibitemOpen
  \bibfield  {author} {\bibinfo {author} {\bibfnamefont {B.}~\bibnamefont
  {Henning}}, \bibinfo {author} {\bibfnamefont {X.}~\bibnamefont {Lu}}, \ and\
  \bibinfo {author} {\bibfnamefont {H.}~\bibnamefont {Murayama}},\ }\href
  {\doibase 10.1007/JHEP01(2016)023} {\bibfield  {journal} {\bibinfo  {journal}
  {JHEP}\ }\textbf {\bibinfo {volume} {01}},\ \bibinfo {pages} {023} (\bibinfo
  {year} {2016})},\ \Eprint {http://arxiv.org/abs/1412.1837} {arXiv:1412.1837
  [hep-ph]} \BibitemShut {NoStop}%
\bibitem [{\citenamefont {Quevillon}\ \emph {et~al.}(2019)\citenamefont
  {Quevillon}, \citenamefont {Smith},\ and\ \citenamefont
  {Touati}}]{Quevillon:2018mfl}%
  \BibitemOpen
  \bibfield  {author} {\bibinfo {author} {\bibfnamefont {J.}~\bibnamefont
  {Quevillon}}, \bibinfo {author} {\bibfnamefont {C.}~\bibnamefont {Smith}}, \
  and\ \bibinfo {author} {\bibfnamefont {S.}~\bibnamefont {Touati}},\ }\href
  {\doibase 10.1103/PhysRevD.99.013003} {\bibfield  {journal} {\bibinfo
  {journal} {Phys. Rev. D}\ }\textbf {\bibinfo {volume} {99}},\ \bibinfo
  {pages} {013003} (\bibinfo {year} {2019})},\ \Eprint
  {http://arxiv.org/abs/1810.06994} {arXiv:1810.06994 [hep-ph]} \BibitemShut
  {NoStop}%
\bibitem [{\citenamefont {Englert}\ \emph {et~al.}(2019)\citenamefont
  {Englert}, \citenamefont {Giudice}, \citenamefont {Greljo},\ and\
  \citenamefont {Mccullough}}]{Englert:2019zmt}%
  \BibitemOpen
  \bibfield  {author} {\bibinfo {author} {\bibfnamefont {C.}~\bibnamefont
  {Englert}}, \bibinfo {author} {\bibfnamefont {G.~F.}\ \bibnamefont
  {Giudice}}, \bibinfo {author} {\bibfnamefont {A.}~\bibnamefont {Greljo}}, \
  and\ \bibinfo {author} {\bibfnamefont {M.}~\bibnamefont {Mccullough}},\
  }\href {\doibase 10.1007/JHEP09(2019)041} {\bibfield  {journal} {\bibinfo
  {journal} {JHEP}\ }\textbf {\bibinfo {volume} {09}},\ \bibinfo {pages} {041}
  (\bibinfo {year} {2019})},\ \Eprint {http://arxiv.org/abs/1903.07725}
  {arXiv:1903.07725 [hep-ph]} \BibitemShut {NoStop}%
\bibitem [{\citenamefont {Alioli}\ \emph {et~al.}(2017)\citenamefont {Alioli},
  \citenamefont {Farina}, \citenamefont {Pappadopulo},\ and\ \citenamefont
  {Ruderman}}]{Alioli:2017jdo}%
  \BibitemOpen
  \bibfield  {author} {\bibinfo {author} {\bibfnamefont {S.}~\bibnamefont
  {Alioli}}, \bibinfo {author} {\bibfnamefont {M.}~\bibnamefont {Farina}},
  \bibinfo {author} {\bibfnamefont {D.}~\bibnamefont {Pappadopulo}}, \ and\
  \bibinfo {author} {\bibfnamefont {J.~T.}\ \bibnamefont {Ruderman}},\ }\href
  {\doibase 10.1007/JHEP07(2017)097} {\bibfield  {journal} {\bibinfo  {journal}
  {JHEP}\ }\textbf {\bibinfo {volume} {07}},\ \bibinfo {pages} {097} (\bibinfo
  {year} {2017})},\ \Eprint {http://arxiv.org/abs/1706.03068} {arXiv:1706.03068
  [hep-ph]} \BibitemShut {NoStop}%
\bibitem [{\citenamefont {Giudice}\ \emph {et~al.}(2007)\citenamefont
  {Giudice}, \citenamefont {Grojean}, \citenamefont {Pomarol},\ and\
  \citenamefont {Rattazzi}}]{Giudice:2007fh}%
  \BibitemOpen
  \bibfield  {author} {\bibinfo {author} {\bibfnamefont {G.~F.}\ \bibnamefont
  {Giudice}}, \bibinfo {author} {\bibfnamefont {C.}~\bibnamefont {Grojean}},
  \bibinfo {author} {\bibfnamefont {A.}~\bibnamefont {Pomarol}}, \ and\
  \bibinfo {author} {\bibfnamefont {R.}~\bibnamefont {Rattazzi}},\ }\href
  {\doibase 10.1088/1126-6708/2007/06/045} {\bibfield  {journal} {\bibinfo
  {journal} {JHEP}\ }\textbf {\bibinfo {volume} {06}},\ \bibinfo {pages} {045}
  (\bibinfo {year} {2007})},\ \Eprint {http://arxiv.org/abs/hep-ph/0703164}
  {arXiv:hep-ph/0703164} \BibitemShut {NoStop}%
\bibitem [{\citenamefont {Liu}\ \emph {et~al.}(2016)\citenamefont {Liu},
  \citenamefont {Pomarol}, \citenamefont {Rattazzi},\ and\ \citenamefont
  {Riva}}]{Liu:2016idz}%
  \BibitemOpen
  \bibfield  {author} {\bibinfo {author} {\bibfnamefont {D.}~\bibnamefont
  {Liu}}, \bibinfo {author} {\bibfnamefont {A.}~\bibnamefont {Pomarol}},
  \bibinfo {author} {\bibfnamefont {R.}~\bibnamefont {Rattazzi}}, \ and\
  \bibinfo {author} {\bibfnamefont {F.}~\bibnamefont {Riva}},\ }\href {\doibase
  10.1007/JHEP11(2016)141} {\bibfield  {journal} {\bibinfo  {journal} {JHEP}\
  }\textbf {\bibinfo {volume} {11}},\ \bibinfo {pages} {141} (\bibinfo {year}
  {2016})},\ \Eprint {http://arxiv.org/abs/1603.03064} {arXiv:1603.03064
  [hep-ph]} \BibitemShut {NoStop}%
\bibitem [{\citenamefont {'t~Hooft}(1999)}]{tHooft:1998qmr}%
  \BibitemOpen
  \bibfield  {author} {\bibinfo {author} {\bibfnamefont {G.}~\bibnamefont
  {'t~Hooft}},\ }\href {\doibase 10.1016/S0920-5632(99)00207-8} {\bibfield
  {journal} {\bibinfo  {journal} {Nucl. Phys. B Proc. Suppl.}\ }\textbf
  {\bibinfo {volume} {74}},\ \bibinfo {pages} {413} (\bibinfo {year} {1999})},\
  \Eprint {http://arxiv.org/abs/hep-th/9808154} {arXiv:hep-th/9808154}
  \BibitemShut {NoStop}%
\bibitem [{\citenamefont {Gross}\ and\ \citenamefont
  {Wilczek}(1973)}]{Gross:1973id}%
  \BibitemOpen
  \bibfield  {author} {\bibinfo {author} {\bibfnamefont {D.~J.}\ \bibnamefont
  {Gross}}\ and\ \bibinfo {author} {\bibfnamefont {F.}~\bibnamefont
  {Wilczek}},\ }\href {\doibase 10.1103/PhysRevLett.30.1343} {\bibfield
  {journal} {\bibinfo  {journal} {Phys. Rev. Lett.}\ }\textbf {\bibinfo
  {volume} {30}},\ \bibinfo {pages} {1343} (\bibinfo {year}
  {1973})}\BibitemShut {NoStop}%
\bibitem [{\citenamefont {Grinstein}\ and\ \citenamefont
  {Randall}(1989)}]{Grinstein:1988wz}%
  \BibitemOpen
  \bibfield  {author} {\bibinfo {author} {\bibfnamefont {B.}~\bibnamefont
  {Grinstein}}\ and\ \bibinfo {author} {\bibfnamefont {L.}~\bibnamefont
  {Randall}},\ }\href {\doibase 10.1016/0370-2693(89)90877-0} {\bibfield
  {journal} {\bibinfo  {journal} {Phys. Lett. B}\ }\textbf {\bibinfo {volume}
  {217}},\ \bibinfo {pages} {335} (\bibinfo {year} {1989})}\BibitemShut
  {NoStop}%
\bibitem [{\citenamefont {Fischler}(1977)}]{Fischler:1977yf}%
  \BibitemOpen
  \bibfield  {author} {\bibinfo {author} {\bibfnamefont {W.}~\bibnamefont
  {Fischler}},\ }\href {\doibase 10.1016/0550-3213(77)90026-8} {\bibfield
  {journal} {\bibinfo  {journal} {Nucl. Phys. B}\ }\textbf {\bibinfo {volume}
  {129}},\ \bibinfo {pages} {157} (\bibinfo {year} {1977})}\BibitemShut
  {NoStop}%
\bibitem [{\citenamefont {de~Florian}\ \emph {et~al.}(2016)\citenamefont
  {de~Florian} \emph {et~al.}}]{LHCHiggsCrossSectionWorkingGroup:2016ypw}%
  \BibitemOpen
  \bibfield  {author} {\bibinfo {author} {\bibfnamefont {D.}~\bibnamefont
  {de~Florian}} \emph {et~al.} (\bibinfo {collaboration} {LHC Higgs Cross
  Section Working Group}),\ }\href {\doibase 10.23731/CYRM-2017-002} {\ \textbf
  {\bibinfo {volume} {2/2017}} (\bibinfo {year} {2016}),\
  10.23731/CYRM-2017-002},\ \Eprint {http://arxiv.org/abs/1610.07922}
  {arXiv:1610.07922 [hep-ph]} \BibitemShut {NoStop}%
\bibitem [{\citenamefont {Froissart}(1961)}]{Froissart:1961ux}%
  \BibitemOpen
  \bibfield  {author} {\bibinfo {author} {\bibfnamefont {M.}~\bibnamefont
  {Froissart}},\ }\href {\doibase 10.1103/PhysRev.123.1053} {\bibfield
  {journal} {\bibinfo  {journal} {Phys. Rev.}\ }\textbf {\bibinfo {volume}
  {123}},\ \bibinfo {pages} {1053} (\bibinfo {year} {1961})}\BibitemShut
  {NoStop}%
\bibitem [{\citenamefont {Alioli}\ \emph {et~al.}(2018)\citenamefont {Alioli},
  \citenamefont {Farina}, \citenamefont {Pappadopulo},\ and\ \citenamefont
  {Ruderman}}]{Alioli:2017nzr}%
  \BibitemOpen
  \bibfield  {author} {\bibinfo {author} {\bibfnamefont {S.}~\bibnamefont
  {Alioli}}, \bibinfo {author} {\bibfnamefont {M.}~\bibnamefont {Farina}},
  \bibinfo {author} {\bibfnamefont {D.}~\bibnamefont {Pappadopulo}}, \ and\
  \bibinfo {author} {\bibfnamefont {J.~T.}\ \bibnamefont {Ruderman}},\ }\href
  {\doibase 10.1103/PhysRevLett.120.101801} {\bibfield  {journal} {\bibinfo
  {journal} {Phys. Rev. Lett.}\ }\textbf {\bibinfo {volume} {120}},\ \bibinfo
  {pages} {101801} (\bibinfo {year} {2018})},\ \Eprint
  {http://arxiv.org/abs/1712.02347} {arXiv:1712.02347 [hep-ph]} \BibitemShut
  {NoStop}%
\bibitem [{\citenamefont {Hammou}\ \emph {et~al.}(2023)\citenamefont {Hammou},
  \citenamefont {Kassabov}, \citenamefont {Madigan}, \citenamefont {Mangano},
  \citenamefont {Mantani}, \citenamefont {Moore}, \citenamefont {Alvarado},\
  and\ \citenamefont {Ubiali}}]{Hammou:2023heg}%
  \BibitemOpen
  \bibfield  {author} {\bibinfo {author} {\bibfnamefont {E.}~\bibnamefont
  {Hammou}}, \bibinfo {author} {\bibfnamefont {Z.}~\bibnamefont {Kassabov}},
  \bibinfo {author} {\bibfnamefont {M.}~\bibnamefont {Madigan}}, \bibinfo
  {author} {\bibfnamefont {M.~L.}\ \bibnamefont {Mangano}}, \bibinfo {author}
  {\bibfnamefont {L.}~\bibnamefont {Mantani}}, \bibinfo {author} {\bibfnamefont
  {J.}~\bibnamefont {Moore}}, \bibinfo {author} {\bibfnamefont {M.~M.}\
  \bibnamefont {Alvarado}}, \ and\ \bibinfo {author} {\bibfnamefont
  {M.}~\bibnamefont {Ubiali}},\ }\href {\doibase 10.1007/JHEP11(2023)090}
  {\bibfield  {journal} {\bibinfo  {journal} {JHEP}\ }\textbf {\bibinfo
  {volume} {11}},\ \bibinfo {pages} {090} (\bibinfo {year} {2023})},\ \Eprint
  {http://arxiv.org/abs/2307.10370} {arXiv:2307.10370 [hep-ph]} \BibitemShut
  {NoStop}%
\bibitem [{\citenamefont {Aad}\ \emph {et~al.}(2019)\citenamefont {Aad} \emph
  {et~al.}}]{ATLAS:2019lsy}%
  \BibitemOpen
  \bibfield  {author} {\bibinfo {author} {\bibfnamefont {G.}~\bibnamefont
  {Aad}} \emph {et~al.} (\bibinfo {collaboration} {ATLAS}),\ }\href {\doibase
  10.1103/PhysRevD.100.052013} {\bibfield  {journal} {\bibinfo  {journal}
  {Phys. Rev. D}\ }\textbf {\bibinfo {volume} {100}},\ \bibinfo {pages}
  {052013} (\bibinfo {year} {2019})},\ \Eprint
  {http://arxiv.org/abs/1906.05609} {arXiv:1906.05609 [hep-ex]} \BibitemShut
  {NoStop}%
\bibitem [{\citenamefont {Sirunyan}\ \emph {et~al.}(2019)\citenamefont
  {Sirunyan} \emph {et~al.}}]{CMS:2018uag}%
  \BibitemOpen
  \bibfield  {author} {\bibinfo {author} {\bibfnamefont {A.~M.}\ \bibnamefont
  {Sirunyan}} \emph {et~al.} (\bibinfo {collaboration} {CMS}),\ }\href
  {\doibase 10.1140/epjc/s10052-019-6909-y} {\bibfield  {journal} {\bibinfo
  {journal} {Eur. Phys. J. C}\ }\textbf {\bibinfo {volume} {79}},\ \bibinfo
  {pages} {421} (\bibinfo {year} {2019})},\ \Eprint
  {http://arxiv.org/abs/1809.10733} {arXiv:1809.10733 [hep-ex]} \BibitemShut
  {NoStop}%
\bibitem [{\citenamefont {de~Blas}\ \emph {et~al.}(2018)\citenamefont
  {de~Blas}, \citenamefont {Criado}, \citenamefont {Perez-Victoria},\ and\
  \citenamefont {Santiago}}]{deBlas:2017xtg}%
  \BibitemOpen
  \bibfield  {author} {\bibinfo {author} {\bibfnamefont {J.}~\bibnamefont
  {de~Blas}}, \bibinfo {author} {\bibfnamefont {J.~C.}\ \bibnamefont {Criado}},
  \bibinfo {author} {\bibfnamefont {M.}~\bibnamefont {Perez-Victoria}}, \ and\
  \bibinfo {author} {\bibfnamefont {J.}~\bibnamefont {Santiago}},\ }\href
  {\doibase 10.1007/JHEP03(2018)109} {\bibfield  {journal} {\bibinfo  {journal}
  {JHEP}\ }\textbf {\bibinfo {volume} {03}},\ \bibinfo {pages} {109} (\bibinfo
  {year} {2018})},\ \Eprint {http://arxiv.org/abs/1711.10391} {arXiv:1711.10391
  [hep-ph]} \BibitemShut {NoStop}%
\end{thebibliography}%

\end{document}